\documentclass[%
 aip,
 amsmath,amssymb,
 reprint,%
]{revtex4-1}

\usepackage{graphicx}
\usepackage{dcolumn}
\usepackage{bm}

\usepackage[utf8]{inputenc}
\usepackage[T1]{fontenc}
\usepackage{mathptmx}

\usepackage[english]{babel}
\usepackage{subfigure}
\usepackage{units}
\usepackage{siunitx} 
\usepackage{amsfonts}
\usepackage{amssymb}
\usepackage{textcomp}          
\usepackage{upgreek}

\begin{document}

\preprint{AIP/123-QED}

\title[MHz repetition rate single-shot electro-optic detection of THz pulses]{Compact single-shot electro-optic detection system for THz pulses with femtosecond time resolution at MHz repetition rates}

\author{B.~Steffen}
\email{bernd.steffen@desy.de}
\affiliation{Deutsches Elektronen-Synchrotron (DESY), Hamburg, Germany}
\author{Ch.~Gerth}
\affiliation{Deutsches Elektronen-Synchrotron (DESY), Hamburg, Germany}
\author{M.~Caselle}
\affiliation{Karlsruhe Institute of Technology (KIT), Eggenstein-Leopoldshafen, Germany}
\author{M. Felber}
\affiliation{Deutsches Elektronen-Synchrotron (DESY), Hamburg, Germany}
\author{T. Kozak}
\affiliation{Deutsches Elektronen-Synchrotron (DESY), Hamburg, Germany}
\author{D.R.~Makowski}
\affiliation{Lodz University of Technology (TUL), \L\'od\'z, Poland}
\author{U. Mavri\v{c}}
\affiliation{Deutsches Elektronen-Synchrotron (DESY), Hamburg, Germany}
\author{A.~Mielczarek}
\affiliation{Lodz University of Technology (TUL), \L\'od\'z, Poland}
\author{P.~Peier}
\altaffiliation[Now at ]{Eidgen\"ossisches Institut f\"ur Metrologie, Bern, Switzerland}
\affiliation{Deutsches Elektronen-Synchrotron (DESY), Hamburg, Germany}
\author{K. Przygoda}
\affiliation{Deutsches Elektronen-Synchrotron (DESY), Hamburg, Germany}
\author{L.~Rota}
\altaffiliation[Now at ]{SLAC National Accelerator Laboratory, Menlo Park, California, United States}
\affiliation{Karlsruhe Institute of Technology (KIT), Eggenstein-Leopoldshafen, Germany}

\date{\today}

\begin{abstract}
Electro-optical detection has proven to be a valuable technique to study temporal profiles of THz pulses with pulse durations down to femtoseconds. 
As the Coulomb field around a relativistic electron bunch resembles the current profile, electro-optical detection can be exploited for non-invasive bunch length measurements at accelerators.
We have developed a very compact and robust electro-optical detection system based on spectral decoding for bunch length monitoring at the European XFEL with single-shot resolution better than \SI{200}{fs}.
Apart from the GaP crystal and the corresponding laser optics at the electron beamline, all components are housed in  \SI{19}{"} chassis for rack mount and remote operation inside the accelerator tunnel.
 An advanced laser synchronization scheme based on radio-frequency down-conversion has been developed for locking a custom-made Yb-fiber laser to the radio-frequency of the European XFEL accelerator.
 In order to cope with the high bunch repetition rate of the superconducting accelerator, a novel linear array detector (KALYPSO) has been employed for spectral measurements of the Yb-fiber laser pulses at frame rates of up to \SI{2.26}{MHz}.  
In this paper, we describe all sub-systems of the electro-optical detection system as well as the measurement procedure in detail, and discuss first measurement results of longitudinal bunch profiles of around 400~fs (rms) with an arrival-time jitter of 35~fs (rms).
\end{abstract}


\maketitle

\section{Introduction}

Electro-optical (EO) detection exploits electromagnetic field-induced birefringence in electro-optically active crystals. 
Probing the induced birefringence with a laser provides a measurement of the temporal profile of fast changing fields.
By combining femtosecond laser pulses with crystals with cut-off frequencies of up to several terahertz, such as GaP (gallium phosphide) or ZnTe (zinc telluride), sampling of electromagnetic pulses with sub-picosecond resolution can be realized.\cite{Wu95, shan00, jiang98, Jamison03}
Among a manifold of applications in the field of laser-based THz spectroscopy\cite{Yasumatsu2012, Riek2015, Sanjuan2018}, EO detection can be applied at an accelerator for temporal profile measurements of electron bunches by probing their Coulomb field.
The Coulomb field resembles the current profile for relativistic electrons and, therefore, EO detection offers the possibility of realizing a non-invasive longitudinal bunch profile monitor with femtosecond single-bunch resolution.\cite{ yan00, wilke02, berden04, cavalieri05, steffen09a, funkner2019}

X-ray Free-Electron Lasers (XFELs) generate X-ray pulses with unprecedented peak brilliance and femtosecond pulse durations, thus enabling the investigation of dynamics in matter on femtosecond time scales.  
The European XFEL (EuXFEL)\cite{altarelli2015} comprises a superconducting linear accelerator that delivers electron bunches at a repetition rate of up to \SI{4.5}{MHz} in trains of up to 2700 bunches every \SI{100}{ms}.
The electron bunches can be distributed between three undulator beamlines where they generate femtosecond laser-like X-ray pulses at tunable wavelengths between \SI{0.05}{nm} and \SI{6}{nm}.
This concept allows simultaneous operation of up to three user experiments which significantly benefit from the large number of generated X-ray pulses.\cite{vagovic2019, gisriel2019, pandey2019}
The generation of the X-ray pulses is based on the self-amplified spontaneous emission (SASE) process which requires short electron bunches with high peak currents of several kA.
This is realized by compressing stepwise the initially several picosecond-long electron bunches created at the photo-cathode gun in three magnetic bunch compressor chicanes between the four accelerating sections.
For the setup and control of the bunch compression process, monitoring of the longitudinal bunch properties, i.e.\ longitudinal bunch profile, beam energy and arrival time, is essential.

Transverse deflecting structures\cite{behrens12, behrens14} provide excellent time resolution for longitudinal bunch profile measurements.
Single electron bunches are streaked, by which the longitudinal coordinate is transformed into a transverse coordinate, and then imaged on a view screen by a 2D camera system.
However, the bunch properties are degraded by the view screen which inhibits the SASE process, and the 2D camera systems cannot cope with the high repetition rate of superconducting accelerators.
In this paper, we present a compact EO detection system designed for non-invasive longitudinal bunch profile monitoring with MHz repetition rates at the EuXFEL.
The EO detection system has been installed inside the accelerator tunnel downstream of the second bunch compression chicane at a beam energy of \SI{700}{MeV}. 
The time resolution of about \SI{150}{fs} matches the design bunch lengths (rms) in the range \SI{200}{fs} to \SI{300}{fs} for bunch charges between \SI{100}{pC} and \SI{1}{nC}, respectively.\cite{Zagorodnov2019}

In Sec.~\ref{sec:EO}, we provide an overview on EO techniques for THz pulse detection. 
In Sec.~\ref{sec:Setup}, we describe the individual sub-systems of the EO detection system including the Yb-fiber laser (Sec.~\ref{sec:Ytterbium_Fiber_Laser}), optics setup with GaP crystal at the accelerator beamline (Sec.~\ref{sec:optics-setup}), novel laser synchronization scheme (Sec.~\ref{sec:synchronization}), and spectrometer with the MHz line detector KALYPSO\cite{rota2019} (Sec.~\ref{sec:spectrometer}).
In Sec.~\ref{sec:Results} we discuss first results of longitudinal bunch profile and bunch arrival-time measurements. Sec.~\ref{sec:conclusion} provides a summary and conclusions..

\section{Electo-optical techniques for {THz} pulse detection}
\label{sec:EO}

The electro-optical effect in GaP or ZnTe is dominated by the Pockels effect and, therewith, depends linearly on the electric field strength of the THz pulse for field strengths of up to MV/m. 
For a laser, passing through an EO crystal, the two orthogonal components of the laser polarization, oriented along the principal axes of the crystal, receive a relative phase shift of
\begin{equation} 
\Gamma=\frac{2 \pi d}{\lambda}(n_1-n_2)=\frac{2 \pi d}{\lambda}n_0^3r_{41}E_\mathrm{THz}\quad, \label{phaseshift} 
\end{equation} 
where $\lambda$ is the laser wavelength, $d$ the thickness of the crystal, and $n_1$ and $n_2$ are the refractive indexes along the principal axes. $n_0$ is the refractive index of the (isotropic) crystal at vanishing electric field, $r_{41}$ its EO coefficient and $E_\mathrm{THz}$ is the applied electric field. 
Using wave plates and a polarizer, this relative phase shift or polarization rotation can be transferred into an amplitude modulation of the transmitted laser light and easily measured with a photodiode or spectrometer. 
Setting the wave plates and polarizer to cross polarization, i.e.\ the non-rotated part of the laser is fully blocked, leads to an amplitude which is proportional to $\Gamma^2$ and therefore to $E_\mathrm{THz}^2$. 
Introducing an additional phase shift by a half-wave plate rotated by an angle $\theta$ in front of the polarizer results in an amplitude modulation which is proportional to $\Gamma$ and $E_\mathrm{THz}$ for $\Gamma\ll\theta$ on top of a constant background.\cite{berndthesis}

\begin{figure}
	\includegraphics[width=0.9\columnwidth]{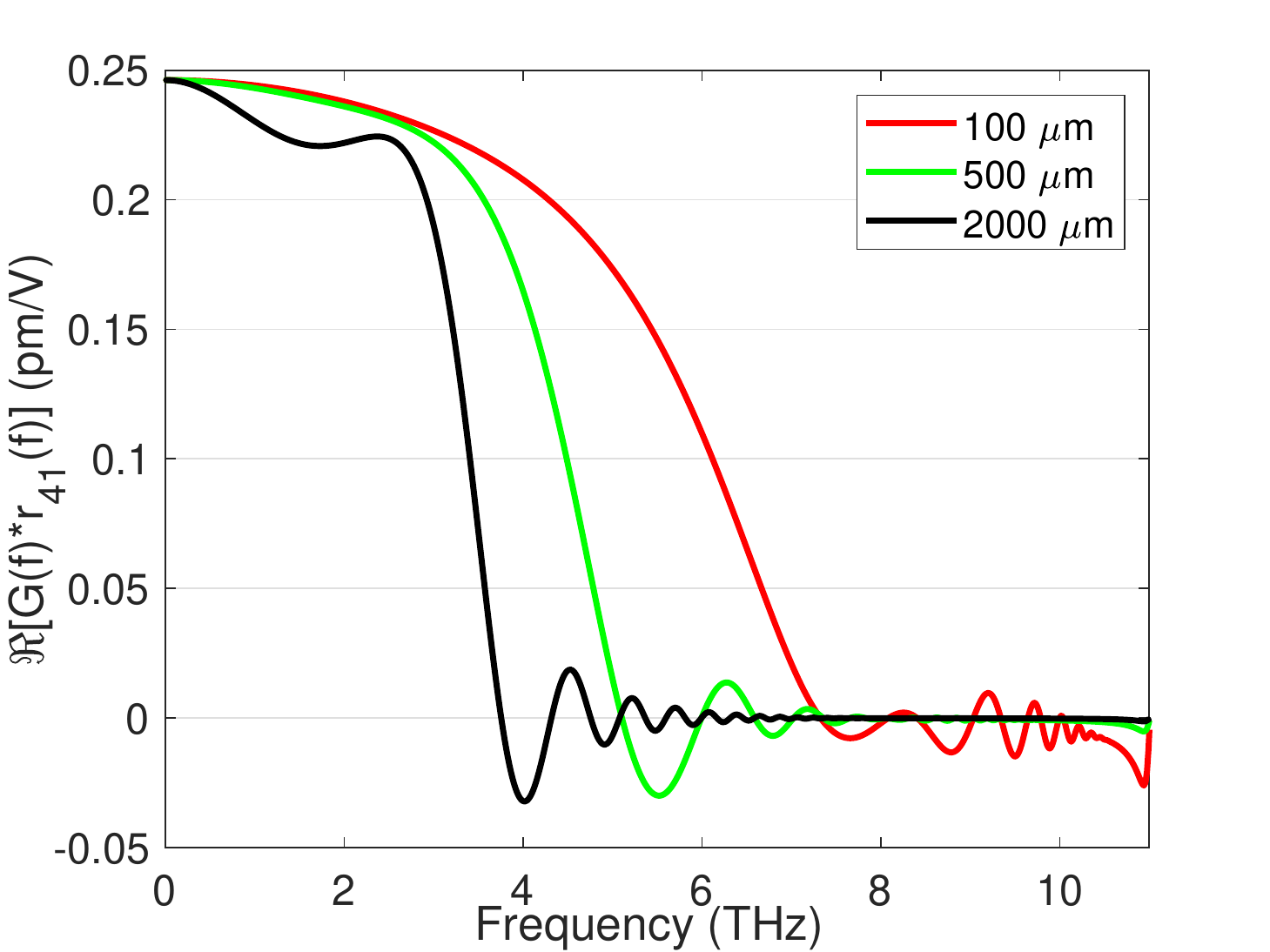}
	\caption{Frequency dependence of  real part of EO efficiency $G(f,d)$ and EO coefficient $r_{41}$ for a GaP crystal at a laser wavelength of \SI{1050}{nm} for three different crystal thicknesses. Calculation based on literature values.\cite{leitenstorfer99}}
  \label{fig_GaPLresp} 
\end{figure}

Phonon resonances in the crystal lead to a frequency dependence of the refractive index $n_0$.  
This results in a frequency-dependent mismatch of the phase velocity of the electric field at THz frequencies and the group velocity of the laser pulse. 
It is customary to introduce an EO efficiency by a complex response function $G(f,d)$, which depends on the frequency of the THz field $f$ and crystal thickness $d$. It includes the effects on the detected field from the frequency-dependent mismatch as well as from the absorption and reflection at the crystal surface. One can now replace the field $E_\mathrm{THz}$ in Eq.~\ref{phaseshift} by an effective detectable field $E_\mathrm{eff}=\Re(r_{41}\cdot G) \cdot E_\mathrm{THz}$ which includes the frequency dependencies of both EO efficiency $G(f,d)$ and EO coefficient $r_{41}$.\cite{wu97, casalbuoni08, steffen09a}
These frequency dependencies limit the range of detectable frequencies and, therewith, the time resolution that can be achieved for the temporal profile of the THz pulse.\cite{paradis2018} 
As an example, the frequency behavior of a GaP crystal has been calculated for three different thicknesses $d$ at a laser wavelength of \SI{1050}{nm} based on literature data\cite{leitenstorfer99} and is depicted in  Fig.~\ref{fig_GaPLresp}. 
With increasing crystal thickness $d$, the upper limit of the detectable frequencies decreases from \SI{8}{THz} for thin GaP crystals below \SI{100}{\upmu m} to about \SI{4}{THz} for a \SI{2}{mm} thick crystal. 
On the other hand, the relative phase shift $\Gamma$ and related thereto the EO signal intensity increase linearly with the crystal thickness.

Aside from EO sampling\cite{Wu95}, where a series of THz pulses is sampled with a short laser pulse by varying the time delay between the laser and THz pulse, other EO detection techniques have been established to realize single-shot resolution.
The temporal profile of a single THz pulse can be encoded either in the transverse profile of a laser pulse (spatially-decoded EO detection)\cite{shan00} or in the spectrum of a chirped laser pulse (spectrally-decoded EO detection).\cite{jiang98} 
Alternatively, the temporal profile of the modulated laser pulse can be measured by a single-shot laser cross-correlator (temporally-resolved EO detection).\cite{Jamison03}

All of the above mentioned techniques have different drawbacks.\cite{berndthesis} 
Temporally-resolved EO detection requires a high laser pulse energy of several hundred \SI{}{\upmu J} to realize a single-shot laser cross-correlator, and any dispersion of the laser pulse after the EO crystal will influence the measured temporal profile and has to be avoided or compensated. 
For spatially-decoded EO detection, the laser has to be focused to a line on the EO crystal and afterward imaged to a line camera, which requires a significant amount of imaging optics and detection electronics close to the electron beamline. 
For the EO detection based on spectral decoding (EOSD), a short (\SI{< 100}{fs}), broadband laser pulse is stretched in a dispersive material or grating stretcher to get a several picosecond-long chirped laser pulse with a known frequency-time relation. 
The part of the laser pulse that overlaps with the THz pulse gets modulated, and the temporal shape of the THz pulse can be retrieved from the modulation of the spectral intensity distribution of the laser pulse.
Frequency mixing between the chirped laser and THz pulse, which leads to signal distortions and becomes severe for strongly stretched laser pulses and short THz pulses, limits the achievable time resolution for the measured temporal profile of the THz pulse.\cite{casalbuoni08}

EOSD was chosen as the best candidate to realize a reliable and robust longitudinal bunch profile monitor inside an accelerator tunnel.
The required laser system can be realized compact, and the laser pulses can be guided to the optics setup at the accelerator beamline and back to the spectrometer via short optical fibers, while dispersion in front of the EO crystal is compensated within the optics setup.
Assuming reasonable parameters for the chirped laser pulse, the resulting signal distortions due to frequency mixing are small enough to achieve a time resolution of about \SI{150}{fs}.


\section{EOSD longitudinal bunch profile monitor} 
\label{sec:Setup}

\begin{figure}
	\includegraphics[width=1.0\columnwidth]{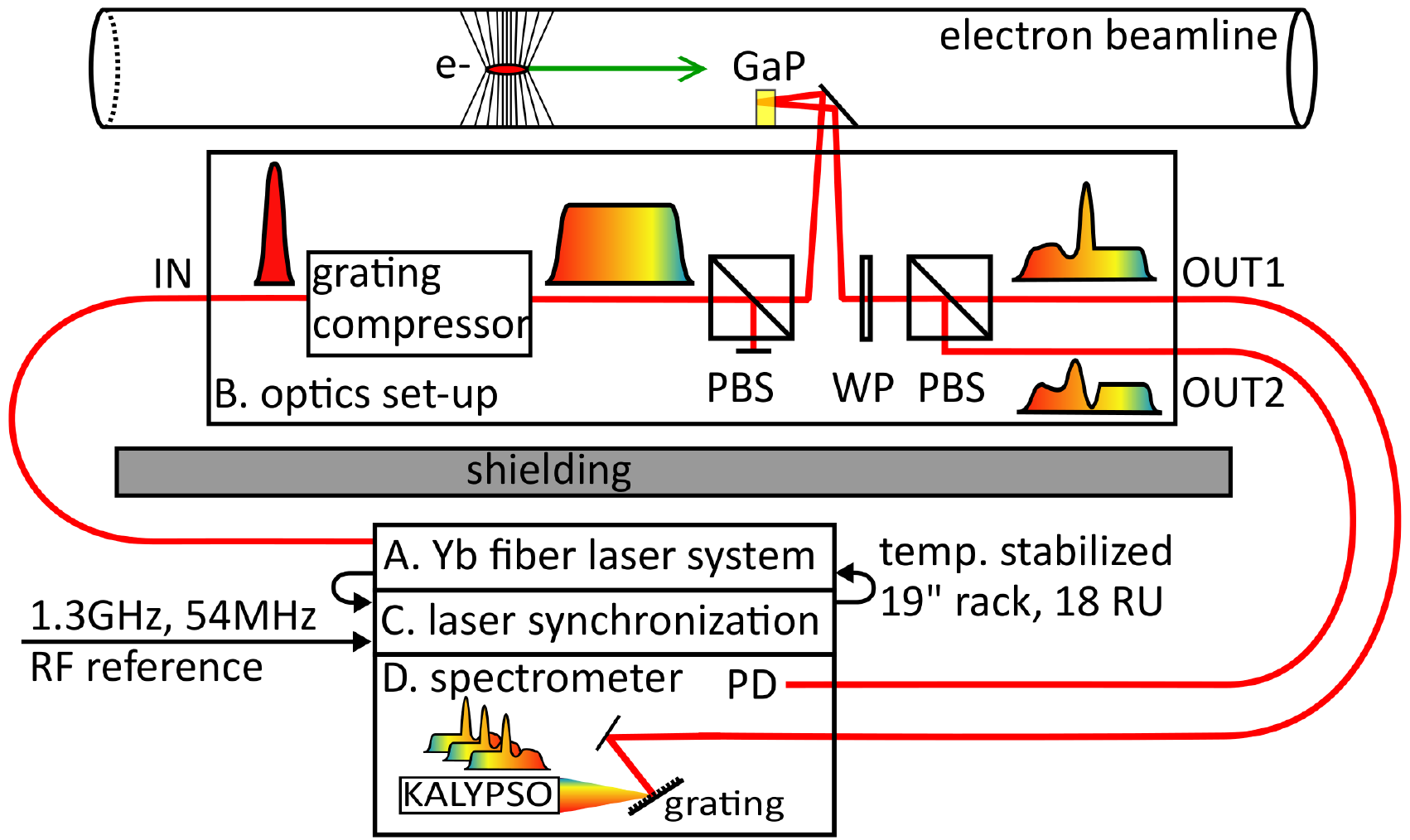}
 	\caption{Scheme of the EOSD setup inside the EuXFEL tunnel. An optics setup is mounted directly at the electron beamline, all other sub-systems are accommodated in a radiation shielded and temperature stabilized rack. (PBS: polarizing beam splitter, WP: quarter- and half-wave plates, PD: photodiode)}
  \label{EOSD_scheme_19inch} 
\end{figure}

In order to achieve a compact and robust design for a diagnostics system to be placed inside an accelerator environment, all sub-systems except the optics setup, which is mounted to the electron beamline and holds the GaP crystal inside the accelerator vacuum, are designed to be housed in  \SI{19}{"} chassis and installed in temperature-stabilized and radiation-shielded racks underneath the electron beamline. 
The total height of all \SI{19}{"} chassis adds up to 18 rack units (RU).
Furthermore, the EOSD setup has been designed to be fully remote controllable as it is inaccessible during accelerator operation.

A schematic layout of the EOSD setup is depicted in Fig.~\ref{EOSD_scheme_19inch}.
It consists of four major sub-systems, which are described in detail in the following sub-sections:
(Sec.~\ref{sec:Ytterbium_Fiber_Laser}) an Yb-fiber laser (\SI{4}{RU}) which is connected via an optical fiber to (Sec.~\ref{sec:optics-setup}) the optics setup at the electron beamline with the supporting control electronics (\SI{4}{RU}); (Sec.~\ref{sec:synchronization}) the laser synchronization to the radio-frequency (RF) reference of the accelerator with analog electronics (\SI{5}{RU}) and fast digital control realized in MicroTCA.4\cite{mtca_picmic}  (\SI{2}{RU}), which is the crate standard at the EuXFEL for the accelerator control; and (Sec.~\ref{sec:spectrometer}) a grating spectrometer (\SI{3}{RU}) which incorporates the linear array detector KALYPSO for spectral measurements of the laser pulses at repetition rates of up to \SI{2.26}{MHz}.

\subsection{Ytterbium fiber laser system}
\label{sec:Ytterbium_Fiber_Laser}

\begin{figure}
	\includegraphics[width=0.9\columnwidth]{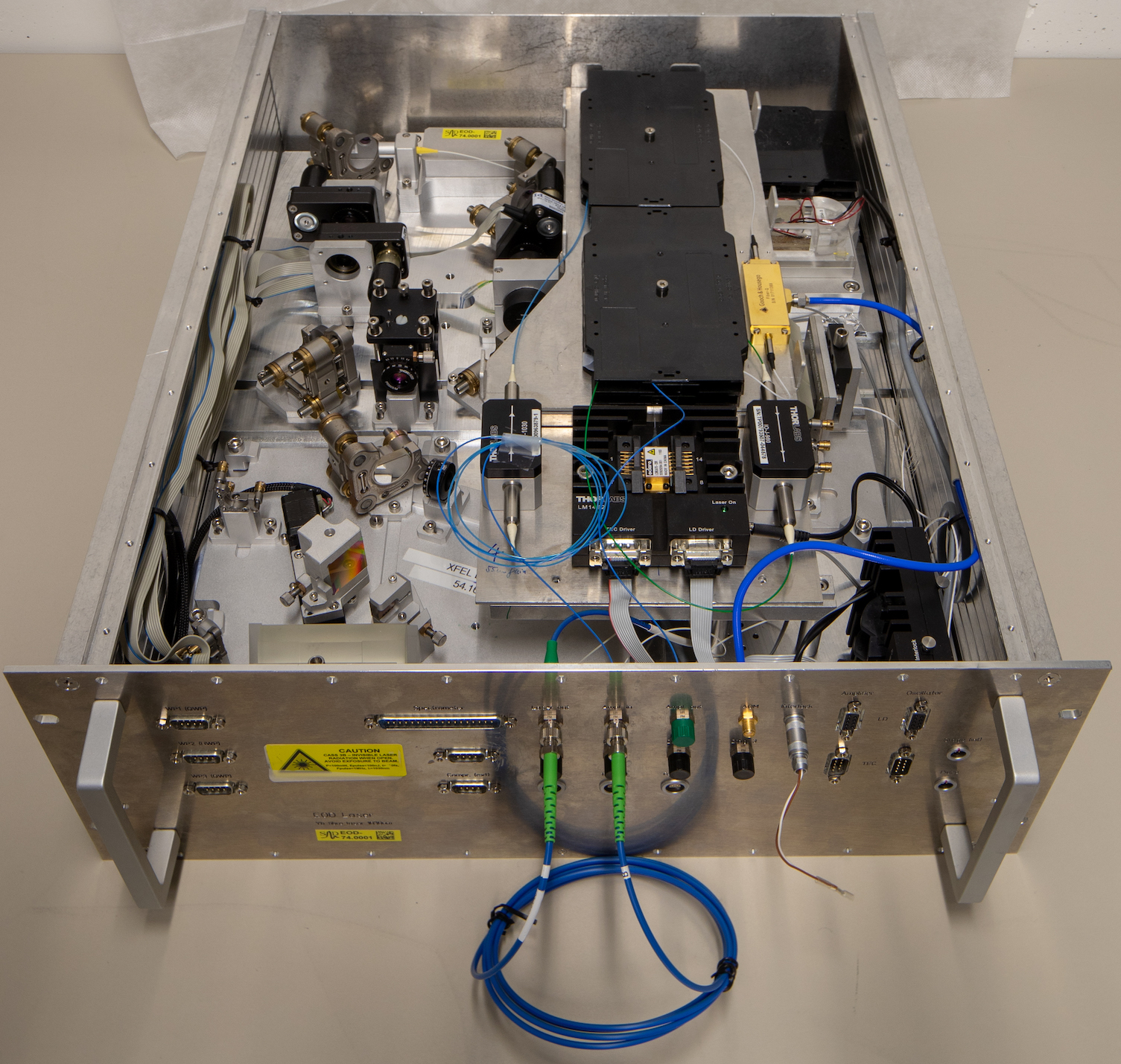}
 	\caption{Photograph of the Yb laser including the oscillator with the fiber and free space part. The single-pass amplifier is mounted on top of the oscillator inside the same housing (photo: D.~N\"olle, DESY).}
  \label{fig:laser_photo} 
\end{figure}

The layout of the laser has been optimized for the generation of short laser pulses with a broad spectral distribution as required for the EOSD technique.
The current version has been adapted from the original design developed at the Paul-Scherrer-Institute (Switzerland)\cite{mueller09, felixthesis} to fit into a \SI{19}{"} chassis with \SI{4}{RU}. 
A photograph of the Yb laser, which consists of an oscillator and single-pass amplifier, is shown in Fig.~\ref{fig:laser_photo}.
The laser oscillator consists of a fiber part including an Yb-doped gain fiber and a free space part which comprises a grating compressor to compensate the dispersion of the fiber and other components for the mode locking mechanism based on non-linear polarization evolution.\cite{Hofer91}
A piezo fiber stretcher is used to synchronize the oscillator to a repetition rate of \SI{54}{MHz}, which is the 24$^{th}$ sub-harmonic of the \SI{1300}{MHz} RF reference signal of the accelerator (see Sec.~\ref{sec:synchronization} for a detailed description). 
The laser oscillator is mounted to a \SI{30}{mm} thick aluminum base plate which is temperature stabilized to a few mK (peak-to-peak) inside the \SI{19}{"} chassis. 

\begin{figure}
	\includegraphics[width=0.9\columnwidth]{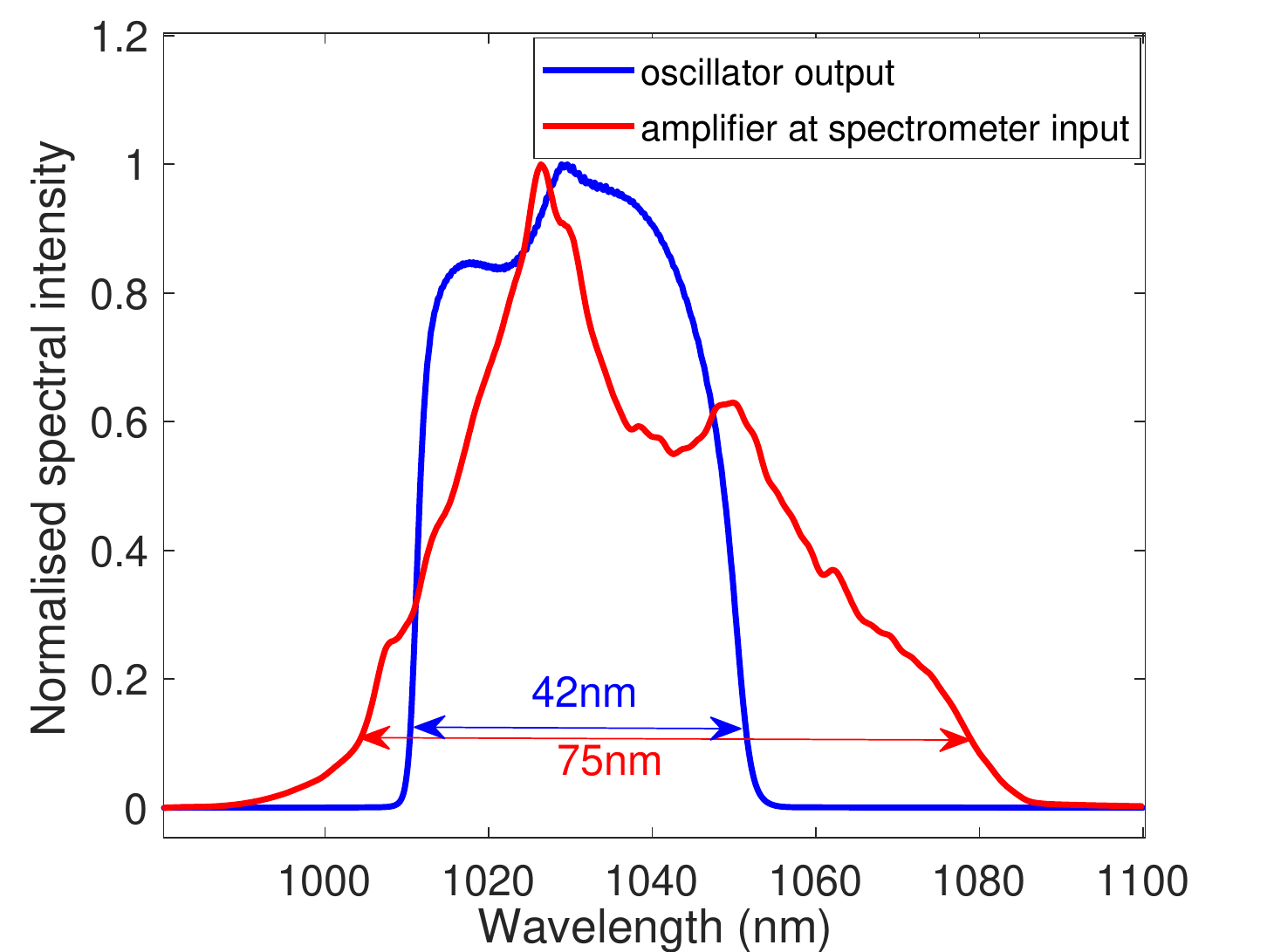}
 	\caption{Spectral distribution of the laser pulses from oscillator (blue) and amplifier (red) measured with an optical spectrum analyzer. The bandwidth at 10\% of the peak is indicated, representing the approximate bandwidth useful for EOSD.}
  \label{fig:laser_spectrum} 
\end{figure}

After the oscillator, a second grating compressor adapts the dispersion in order to optimize the pulse width for the following fiber amplifier.
An acusto-optical modulator (AOM) is used to generate a replica of the electron bunch pattern, i.e.\ millisecond-long bursts at \SI{1.13}{MHz} or \SI{2.26}{MHz} with a repetition rate of \SI{10}{Hz}, which allows to boost the pulse energy in the amplifier without increasing the average power of the pump laser diode. The AOM is triggered from a commercially available MicroTCA.4-compliant board (NAMC-psTimer, N.A.T.) of the EuXFEL timing system\cite{x2Timer}, which is used for the distribution of the accelerator clock and trigger signals to every crate via optical links. Clock and trigger signals are distributed inside each MicroTCA crate via dedicated timing lines of the backplane to other electronic boards or via RJ-45 sockets to external devices as the AOM.

For EOSD, the spectral width directly determines both the resolution and temporal range of the measurement\cite{jiang98, wilke02}.
Due to the large product of bandwidth, dispersion and interaction length in the fiber part, both the oscillator and amplifier produce chirped pulses with a length of up to \SI{10}{ps} (FWHM). 
The spectral phase of the oscillator is mainly linear and, therefore, the pulse length can easily be compressed with grating compressors down to about \SI{60}{fs} (FWHM). 
If properly pre-chirped, the amplified pulses become short and the power density increases near the end of the fiber which leads to strong non-linear effects and a broadening of the spectrum. 
The broadened spectrum shifts from \SI{1030}{nm} towards \SI{1050}{nm} and provides a bandwidth of almost \SI{80}{nm} which can be used for EOSD measurement. 
Typical spectral distributions of the laser pulses recorded with an optical spectrum analyzer after the oscillator and amplifier are depicted in Fig.~\ref{fig:laser_spectrum}, and the usable bandwidths are indicated.   
After the amplifier and non-linear broadening of the spectrum, the spectral phase has also lager contributions of higher order terms which complicates the pulse compression. 
Nevertheless, the pulses can still be compressed with grating compressors to less than \SI{50}{fs} (FWHM) but with some remaining pedestals. 
The parameters of the laser are summarized in Table~\ref{laserspec}. 
\begin{table}[tbh]
  \centering
  \caption{Parameters of the Yb-fiber laser system.}
    \begin{tabular}{lll}
      \hline
      \textbf{} & \textbf{Oscillator} & \textbf{Amplifier} \\
      \hline
      \hline
      Average power &  20 -- 50~mW & 5 -- 100~mW \\
      Repetition rate & 54~MHz & 0.1 -- 2~MHz \\
      Pulse length (compr.) & \SI{60}{fs} (FWHM) & \SI{50}{fs} (FWHM) \\
      Pulse energy & 0.4 -- 1~nJ & 5 -- 100~nJ\\
      Usable bandwidth & \SI{40}{nm} & 40 -- 80~nm \\
      Central wavelength & \SI{1030}{nm} & 1030 -- 1050~nm \\
      \hline
    \end{tabular}
  \label{laserspec}
\end{table}

The laser pulses, for which the pulse energy can be adjusted between \SI{5}{nJ} and \SI{100}{nJ}, are transported via a \SI{6}{m}-long polarization maintaining optical fiber to the optics setup at the accelerator beamline. 
Had the laser system been installed outside the EuXFEL accelerator tunnel, the fiber length would have have been several \SI{100}{m}. 
This would have introduced second- and third-order dispersion which could not have been compensated by a grating compressor alone,\cite{azima06} and the long fiber would have added jitter and drift to the arrival time of the laser pulses at the optics setup.

\begin{figure}
   \centering
   \includegraphics[width=.9\columnwidth]{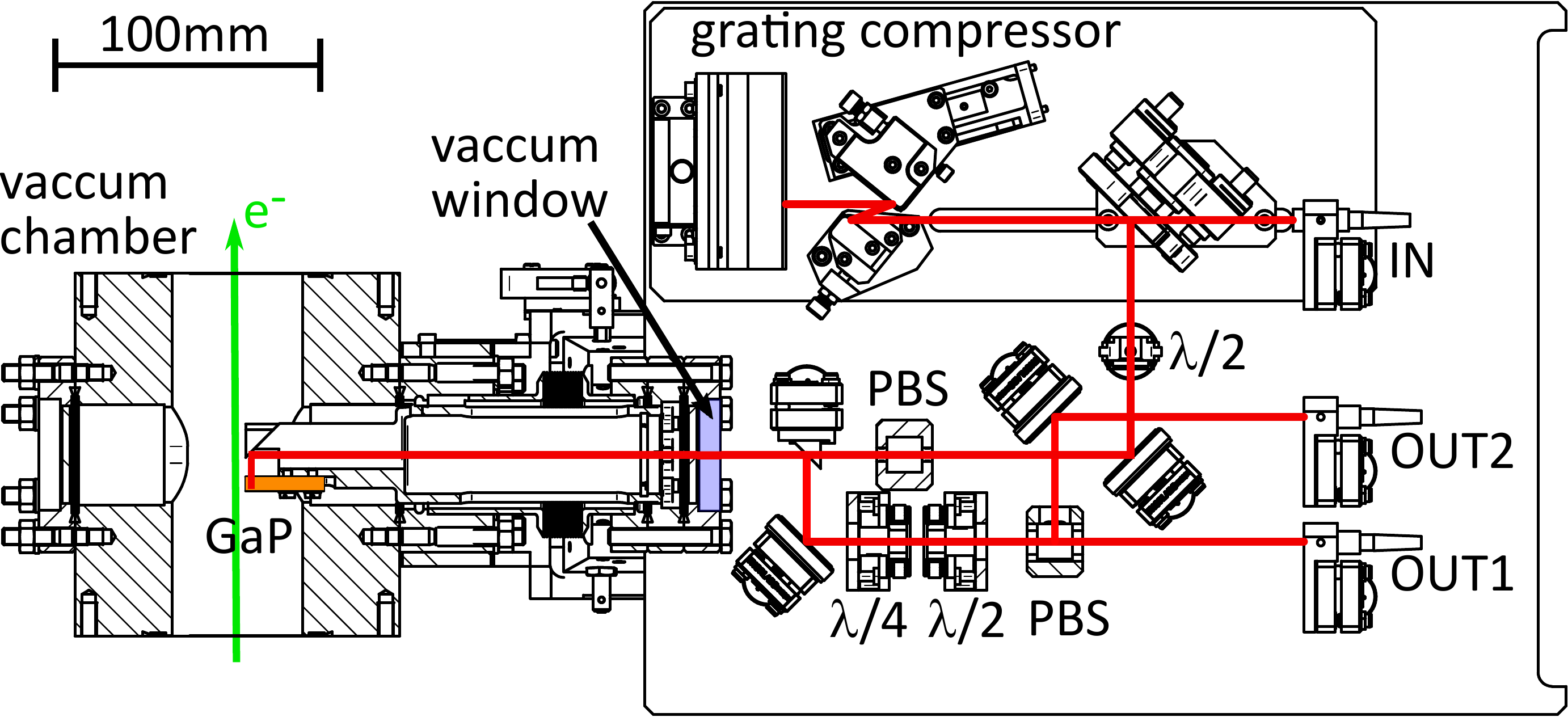}
   \caption{Assembly drawing of the optics setup including the vacuum feed-through and vacuum chamber at the electron beamline. IN: fiber from laser; OUT1/2: fibers to spectrometer and photodiode; PBS: polarizing beam splitter; $\lambda/2$ and $\lambda/4$: wave plates; GaP: gallium-phosphite crystal.}
   \label{fig:Opticbox}
\end{figure}

\subsection{Optics setup at the electron beamline}
\label{sec:optics-setup}

A schematic drawing of the optics setup mounted to the vacuum chamber at the electron beamline is shown in Fig.~\ref{fig:Opticbox}.
The GaP crystal and one mirror are placed inside the electron beamline on a holder that is mounted on a motorized linear motion vacuum feed-through. 
The feed-through also comprises a fused-silica window with an anti-reflection coating for \SI{1050}{nm} and an engineered baseplate (\SI{340}{mm} $\times$ \SI{240}{mm}) which is rigidly fixed to the feed-through and holds all optical elements including the fiber couplers. 
In this way, the position of the GaP crystal relative to all optical elements between the fiber coming from the laser and the fibers going to the spectrometer is fixed, avoiding any misalignment or timing changes when the linear stage is moved to adjust the crystal position with respect to the electron beam.

The first optical element after the fiber coupler connected to the laser is a grating compressor which is used to adjust the length and chirp of the laser pulses. 
The compressor consists of two gratings with \SI{1200}{lines/mm}, one of which is mounted on a motorized stage, to be able to adjust the pulse length from full compression to more than \SI{5}{ps} (FWHM). 
The laser pulses are guided by silver coated mirrors under a small vertical angle inside the vacuum chamber to the GaP crystal (thickness \mbox{$d$ = \SI{2}{mm}}) via an achromatic half-wave plate and polarizer to optimize transmission and polarization. 
The laser pulses first counter-propagate with respect to the electron bunches, enter the GaP crystal on the anti-reflective coated rear side, and are then reflected from the highly-reflective coated front side. 
The reflected laser pulses now co-propagate with the Coulomb field of the electron bunches which induce a modulation of laser polarization inside the crystal. 
The laser pulses are reflected back through the vacuum window and pass through quarter-wave and half-wave plates to optimize the polarization for the measurement, and are then split into their orthogonal polarization components using a polarizing beam splitter. 
Both polarizations are coupled into fibers (denoted OUT1 and OUT2 in Fig.~\ref{fig:Opticbox}), which are connected to a spectrometer for the measurement of the spectral distribution of the laser pulses, and a photo-diode for coarse timing detection. 
The spectrometer and read-out electronics may be placed outside the accelerator tunnel; however, this would require an additional MicroTCA.4 crate for the read-out and introduce additional latency (\SI{5}{\upmu s/km}) and attenuation (\SI{1.5}{dB/km}) to the laser pulses. 
A detailed description of the spectrometer is given in Sec.~\ref{sec:spectrometer}.

\subsection{Laser synchronization to RF reference}
\label{sec:synchronization}

The Yb-fiber laser is synchronized to the RF reference of the EuXFEL accelerator with the help of a motorized optical delay stage (for larger corrections of the repetition rate) and a piezo fiber stretcher (for small but fast changes) inside the laser oscillator.  
By changing the oscillator cavity length, the repetition rate and relative timing of the laser pulses with respect to the electron bunches can be adjusted. 
One of the key components for a synchronization setup is the phase detector, which measures the phase difference or, in the unsynchronized state, frequency offset between the laser pulse train and RF reference. 
In this paper, we present an advanced phase detection scheme based on RF down-conversion to an intermediate frequency (IF) which offers in comparison to direct RF-sampling increased sensitivity of the phase detector. 
A field-programmable-gate-array (FPGA) is used to implement a fully digital controller which enables implementation of complex algorithms and access to various intermediate signals within the control loop. 
Other advantages of this scheme are that detection of high-frequency signals can be avoided and a vector-modulator is not required as phase shifter for timing scans. 

The underlying principle of the down-conversion scheme is based on the proper choice of a fundamental laser repetition rate, which needs to be the $n^\mathrm{th}$ subharmonic of the RF reference.
The $(n+1)^\mathrm{th}$ harmonic of the laser is then down-converted with the RF reference, and the frequency of the resulting sine-wave is the difference of the two frequencies, which is exactly the fundamental laser repetition rate in the case of a synchronized laser. 
An offset in the laser repetition rate with respect to the synchronized case results in $(n+1)$-times the frequency offset in the down-converted signal. 
For a synchronized laser, the phase of the output signal will shift, if a phase shift is introduced between the laser and the RF reference, by $(n+1)$-times the introduced phase shift.

\begin{figure}
	\includegraphics[width=0.8\columnwidth]{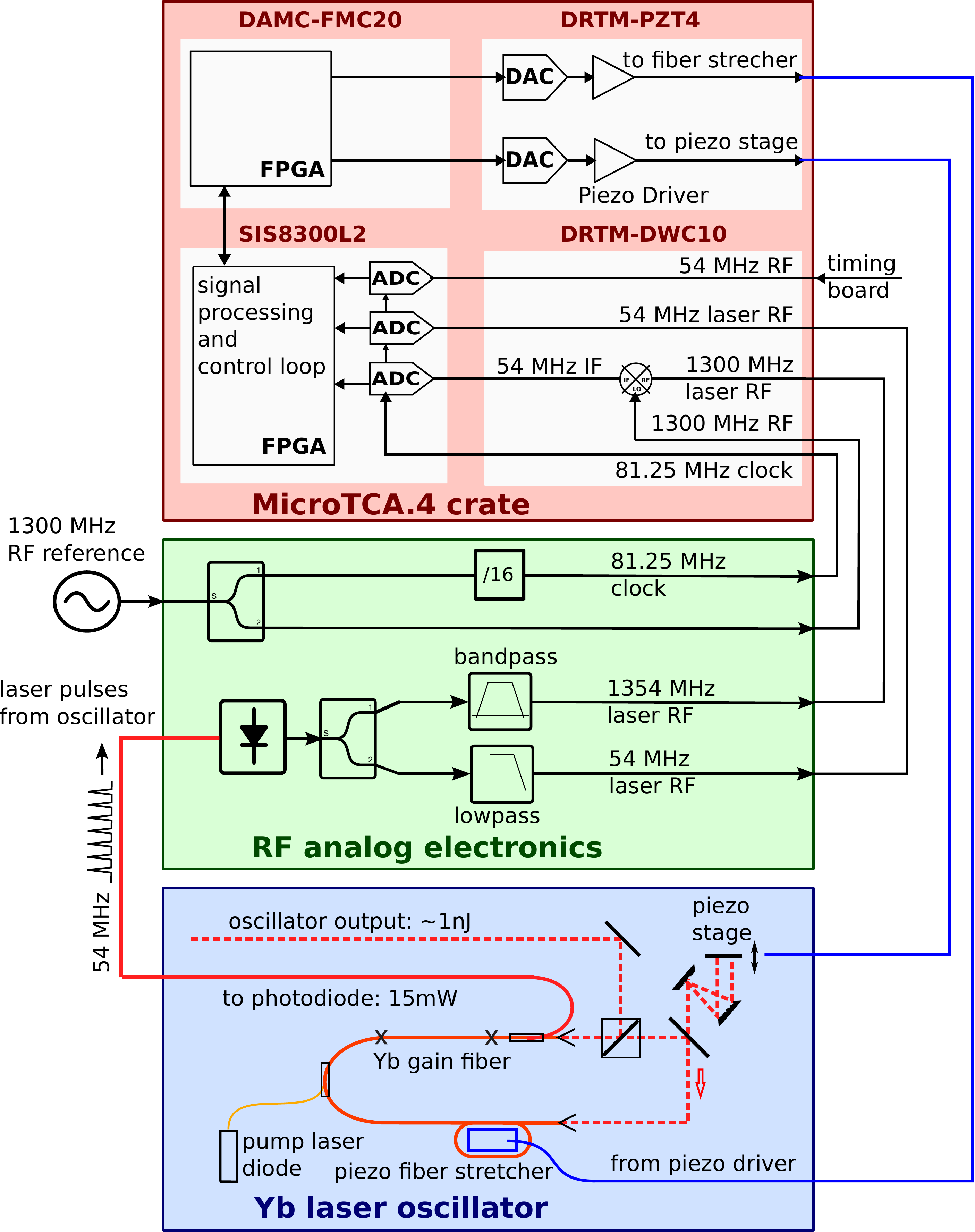}
 	\caption{Block diagram of the laser synchronization to the RF reference including the Yb laser oscillator, RF analog electronics, and MicroTCA.4 components.}
  \label{fig:Scheme_EOD_Synch_simple} 
\end{figure}

Figure~\ref{fig:Scheme_EOD_Synch_simple} shows a block diagram of the Laser to RF synchronization. 
To generate RF signals from the laser, an optical signal (\SI{15}{mW}) is tapped off from the laser pulses in the Yb laser oscillator and fed via fiber to the RF analog electronics in a \SI{19}{''} chassis with 4 RU.  
A photodiode together with an RF-splitter and band-pass filters are used to convert the laser pulses into two RF signals at \SI{54}{MHz} and \SI{1354}{MHz}, which are the fundamental and 25$^{\rm th}$ harmonic of the laser repetition rate, respectively.
The master oscillator of the EuXFEL  provides the RF reference of \SI{1300}{MHz} to which the laser has to be synchronized. 
The clock signal of \SI{81}{MHz} for the analog-to-digital conversion (ADC) of the down-converted signal is generated from the RF reference with a low-noise divider to avoid clock timing jitter.

The laser RF (\SI{1354}{MHz}), RF reference (\SI{1300}{MHz}), and ADC clock signal (\SI{81}{MHz}) are fed from the analog RF analog electronics to a down-converter board (DRTM-DWC10, Struck Innovative Systeme GmbH) which is placed as rear transition module (RTM) inside the MicroTCA.4 crate.
The down-converted IF of \SI{54}{MHz} and ADC clock signal (\SI{81}{MHz}) are directly transmitted to a 10-channel \SI{125}{MSPS} digitizer board (SIS8300L2, Struck Innovative Systeme GmbH) located in the front-side of the crate. 
The digital signal processing in the FPGA of the digitizer board includes non-I/Q detection\cite{doolittle2006} of the sampled signals, amplitude and phase transformation, filtering, feedback controller, and feed-forward tables. 
The output of the controller is transmitted over low-latency links on the backplane of the MicroTCA.4 crate to the neighboring board (DAMC-FMC20, CAENels), which is a FPGA mezzanine card (FMC) carrier that
handles the data transmission to the MicroTCA.4 compliant piezo driver board (DRTM-PZT4, Piezotechnics GmbH).\cite{przygoda2017} 
This RTM can drive up to four piezos in parallel with capacitances of up to few $\upmu\mathrm{F}$. 
In this application, two outputs are used to drive the piezo fiber stretcher and piezo stepper motor stage inside the Yb laser oscillator for fine and coarse tuning of the laser cavity length, respectively.

In order to detect the timing offset of the laser pulses with respect to the electron bunches, both the \SI{54}{MHz} accelerator-synchronous reference signal, generated by the timing card inside the MicroTCA.4 crate, and laser RF of \SI{54}{MHz} are digitized by two ADCs directly, and their phase difference is monitored.
In a first step, the laser is synchronized using the \SI{54}{MHz} Laser RF for setting the absolute timing on a picosecond level and afterwards the signal input of the synchronization control loop is swapped to the down-converted \SI{1354}{MHz} signal as the sensitivity to phase errors is 25 times larger here. 
To scan the timing offset of the laser pulses, the phase set-point of the controller can be varied infinitely with a speed of up to \SI{200}{ps/s}, depending on the gain of the control loop of the synchronization.

\begin{figure}
	\includegraphics[width=0.95\columnwidth]{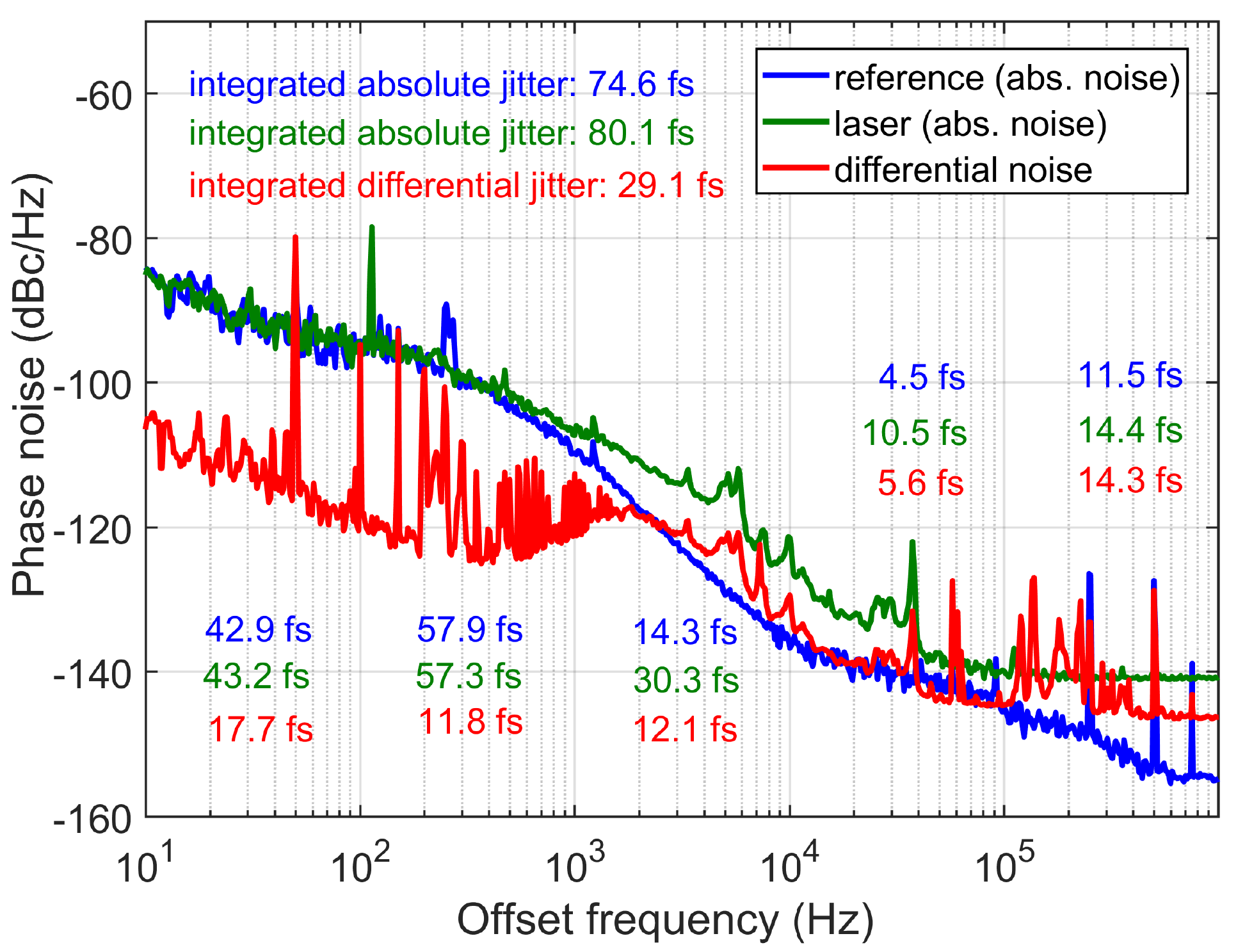}
 	\caption{Phase noise spectra of the 25$^{\rm th}$ harmonic (\SI{1354}{MHz}) of the fundamental repetition rate (\SI{54}{MHz}) of the locked laser (green curve), of the RF reference at \SI{1300}{MHz} (blue curve), and the baseband noise of the 24$^{\rm th}$ harmonic (\SI{1300}{MHz}) of the laser mixed with the reference (red curve). The rms jitter per decade is indicated with colored numbers.}
  \label{fig:phase_noise}
\end{figure}

The performance of the laser synchronization is limited by the phase noise of the synchronized laser at the locking frequency of \SI{1300}{MHz}, from which the remaining integrated timing jitter can be calculated. 
Figure~\ref{fig:phase_noise} shows the in-loop phase-noise of the laser RF (green curve) and RF reference (blue curve) measured in a laboratory environment. 
It can be seen that the locking to the RF reference has strong influence to the laser up to \SI{1}{kHz} offset frequency, i.e.\ the phase noise curve of the laser RF follows the phase noise curve of the RF reference for lower frequencies. 
Amplified noise and resonances in the feedback control loop prevent higher locking bandwidths.

The baseband noise (red curve in Fig.~\ref{fig:phase_noise}) shows the differential timing jitter between the laser and RF reference measured independently (out-of-loop) with a \SI{1300}{MHz} double-balanced mixer at baseband.
Several peaks can be identified, the most prominent is located at \SI{37}{kHz} and corresponds to the piezo fiber-stretcher resonance. 
The plateau at higher offset frequencies (\SI{> 50}{kHz}) is caused by the limited noise performance of the photo receiver. 
The spurious peaks at multiples of \SI{50}{Hz} result from ground loops in the measurement setup and contribute considerably to the integrated jitter.

The total integrated timing jitter (rms) of the RF reference and un-synchronized laser RF in the interval between \SI{10}{Hz} and \SI{1}{MHz} were determined to be \SI{75}{fs} and \SI{80}{fs}, respectively. 
The integrated differential jitter (rms) of the synchronized laser to the RF reference (baseband) amounts to \SI{29}{fs}.

\subsection{Grating spectrometer with MHz linear array detector KALYPSO} 
\label{sec:spectrometer}

For the measurement of the spectral distribution of the chirped laser pulses, one polarization component is sent via an optical fiber to a custom-made grating spectrometer (see  Fig.~\ref{EOSD_scheme_19inch}).
The spectrometer is built from standard, commercially available optics and housed in a \SI{19}{"} chassis with \SI{3}{RU}.
A fiber-coupled collimator illuminates a \SI{2}{"} grating with \SI{600}{lines/mm}, and the first order of the diffracted light is focused by two cylindrical lenses onto the InGaAs microstrip sensor of the linear array detector KALYPSO,\cite{rota2019} which has been designed for EOSD at EuXFEL and the storage ring of the Karlsruhe Research Accelerator (KARA)\cite{funkner2019} with continuous data read-out at frame rates of up to \SI{2.7}{MHz}. 

\begin{figure}
	\includegraphics[width=0.8\columnwidth]{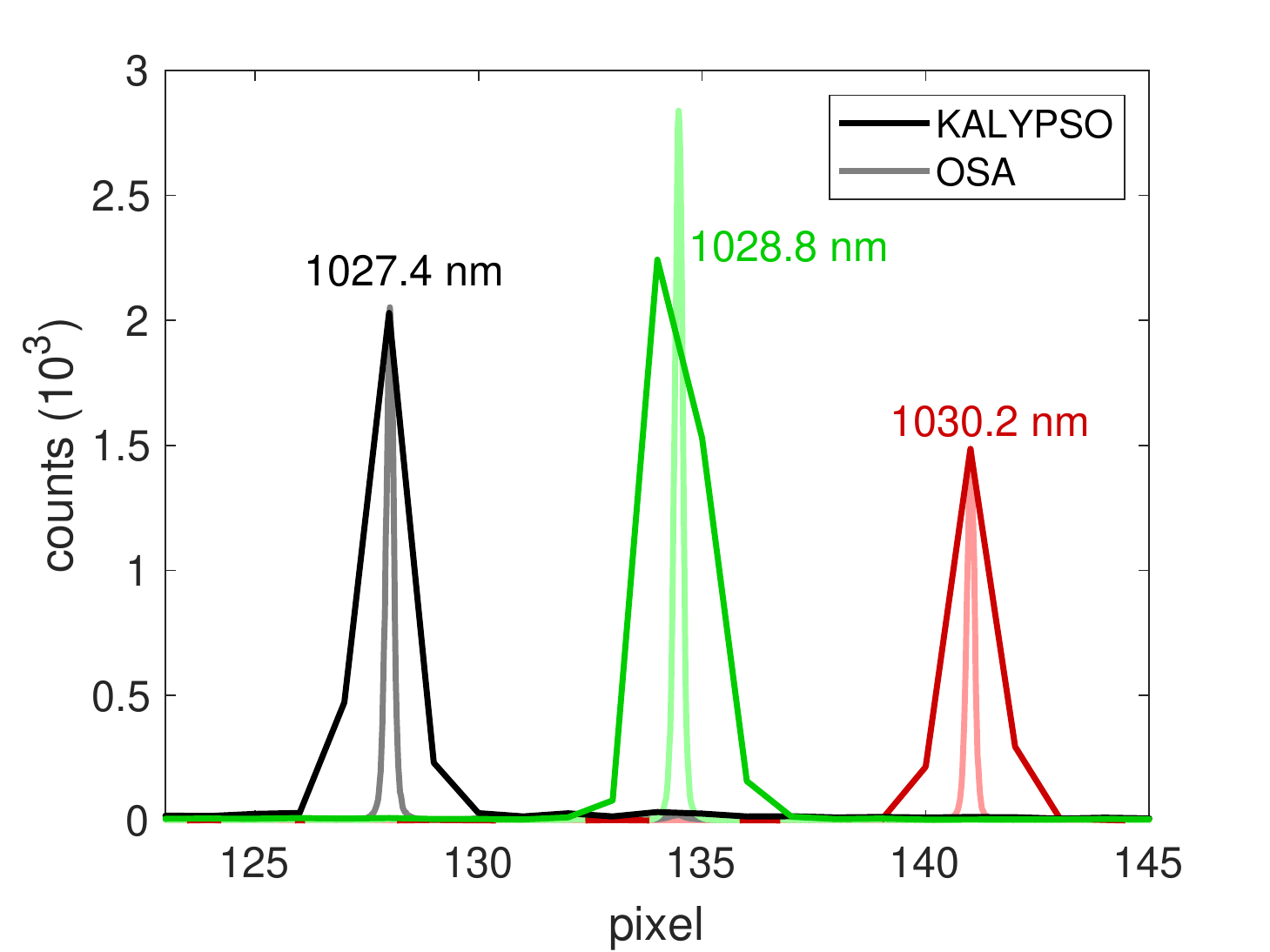}
 	\caption{Spectra of three continuous-wave laser states (mean of 100 spectra each with background subtracted), measured with the spectrometer and the KALYPSO detector. In lighter color the same laser states are shown measured with an optical spectrum analyzer (OSA) with a resolution of \SI{0.02}{nm}.}
  \label{fig:cwspectrum} 
\end{figure}

The spectral resolution was designed to match the pixel size of \SI{50}{\upmu m} of the KALYPSO line array and to cover about \SI{0.2}{nm/pixel}. 
To verify this, Fig.~\ref{fig:cwspectrum} shows three spectra, measured with the Yb-fiber laser operating in continuous-wave mode, which have a bandwidth of less than \SI{0.05}{nm}. 
However, a calibration in wavelength is not necessary as pixels can be calibrated directly to time (see~\ref{sec:meas-proc}).
The achievable time resolution of the EOSD setup is dominated by the phase matching of the THz field of the electron bunches and laser pulses in the GaP crystal (see Sec.~\ref{sec:EO}). 
In comparison, the resolution of the grating in combination with the finite pixel width of the KALYPSO detector is negligible.   

The detector system\cite{gerth19} consists of three main components: (i) The KALYPSO mezzanine card with the radiation sensor and ADC, which is mounted on (ii) the FMC carrier board for data acquisition from the ADC and data transmission to (iii) a MicroTCA.4 board for integration into the EuXFEL accelerator control system.
The radiation sensor on the KALYPSO mezzanine card is a commercially available InGaAs microstrip sensor (ELC-002256, Xenics) which is sensitive in the wavelength range \SI{0.9}{\upmu m} to \SI{1.7}{\upmu m}.
The current version has a total of 256 microstrips with a channel pitch of \SI{50}{\upmu m} and a height of \SI{500}{\upmu m}.
Each microstrip of the InGaAs sensor is bonded to an input channel of two pieces of a modified version of the GOTTHARD chip\cite{mozzanica12} that comprise analog signal amplification and 16:1 multiplexers.
The resulting 16 differential outputs of both GOTTHARD chips are routed 16-channel ADC (AD9249, Analog Devices) with 14-bit resolution that is operated at a sampling rate of \SI{54}{MHz}. 
The number of effective bits that can be used of the 14-bit ADC amounts to 12.7~bits, i.e.\ a range of about 6800 ADC counts.
Compared to the original GOTTHARD chip, the correlated-double-sampling stage and automatic gain switching mechanism have been omitted in order to achieve a maximum frame rate of \SI{2.7}{MHz}.

The FMC carrier board incorporates a FPGA for data acquisition from the ADC on the KALYPSO mezzanine card and fast data processing.
The data transmission and control of the FMC carrier can be realized via four optical links (each providing a data rate of up to \SI{6.5}{Gbps}) with a commercially available MicroTCA.4-compliant board (MFMC, AIES) equipped with a FMC board (FMC-2SFP+, CAENels) for fast SFP communication. 
For this application, one optical link is sufficient, and the total latency for real-time data acquisition and processing of one frame with 256 pixels is less than \SI{1}{\upmu s}.  
Clock and trigger signals for electron bunch synchronous data acquisition of the laser pulse spectra are provided by the NAMC-psTimer card. 
The clock signal is cleaned from jitter in a phase-locked loop on the FMC carrier board and then provided to the FPGA as well as the two GOTTHARD chips and ADC on the KALYPSO mezzanine card.

\section{Experimental results}
\label{sec:Results}

All measurements presented in this section have been performed parasitically during X-ray photon delivery to user experiments or for the commissioning of photon beamlines under lasing conditions of the EuXFEL.  
The data was taken downstream of the second bunch compressor at electron beam energies of \SI{700}{MeV} and bunch charges of \SI{480}{pC} and \SI{240}{pC}, respectively. 
Measurements with single-bunch resolution for every bunch in the bunch train are possible at bunch repetition rates of \SI{1.13}{MHz} or \SI{2.26}{MHz}, whereas for accelerator operation at the maximum repetition rate of \SI{4.5}{MHz} only every second bunch can be measured. 

\subsection{Measurement procedure and time calibration}
\label{sec:meas-proc}

\begin{figure}
 \includegraphics[width=1\columnwidth]{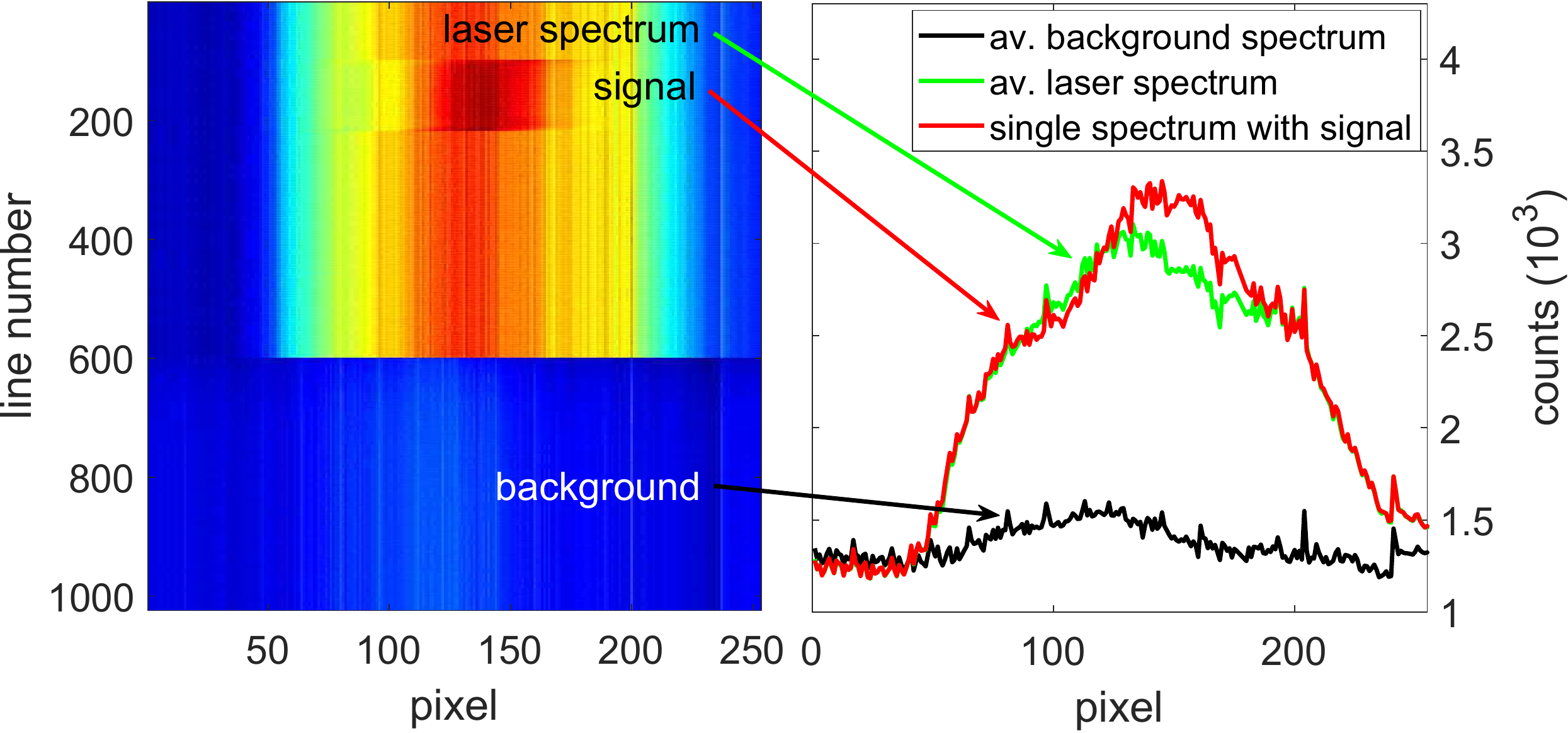}
  \caption{1000 consecutive KALYPSO lines recorded with 1.13~MHz around one electron bunch train at the EuXFEL (left). The averaged background signal, the averaged unmodulated laser spectrum and one laser spectrum modulated by an electron bunch (right).}
  \label{fig:normalizing_EOD1}
\end{figure}

\begin{figure}
 \includegraphics[width=1\columnwidth]{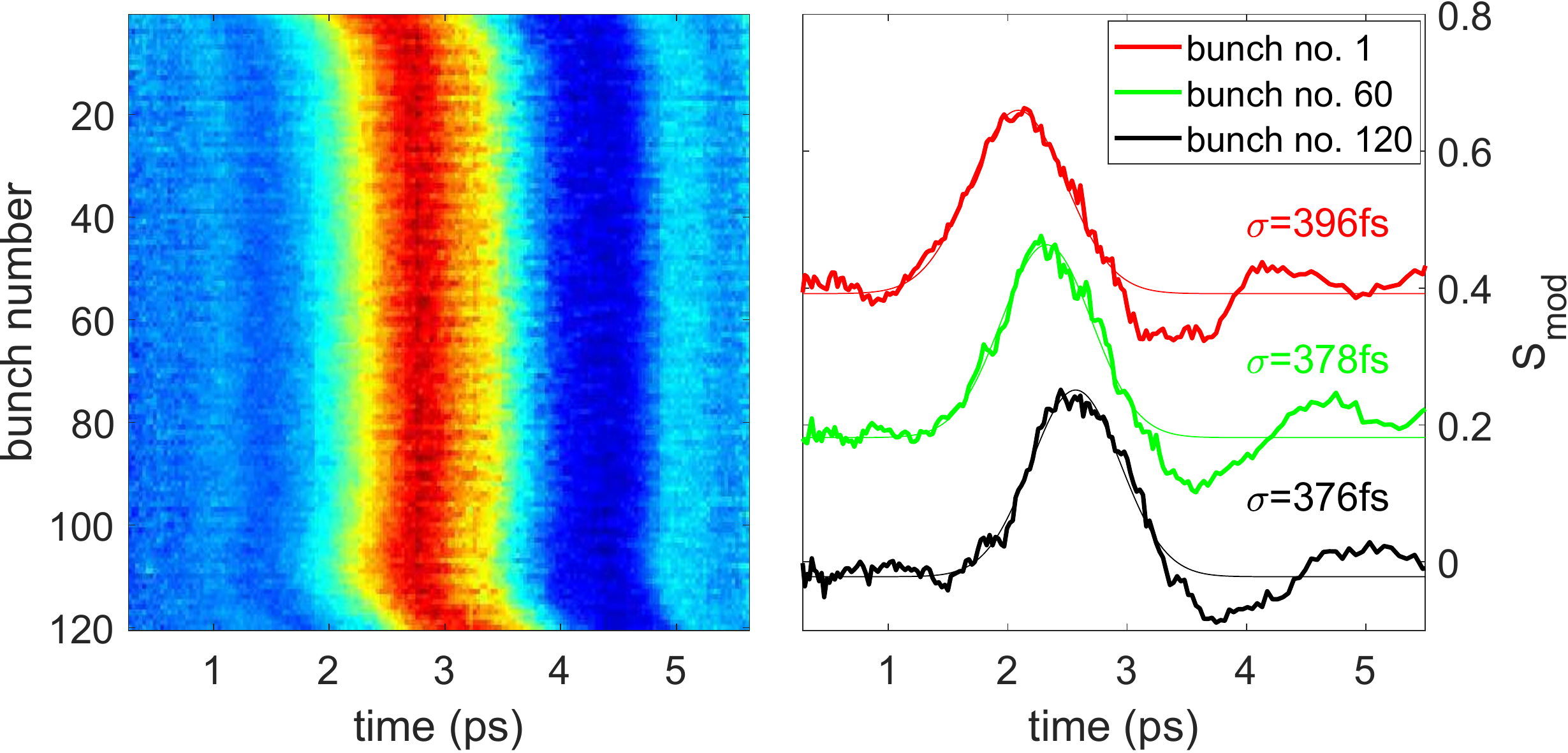}
  \caption{The calculated normalized signal modulation from the 120 electron bunches of the data-set of Fig.~\ref{fig:normalizing_EOD1} (left), and normalized signal traces of three individual bunches out of the bunch train together with Gaussian fits to the longitudinal profiles (right). The traces have been off-set by 0.2 for better visibility.}
  \label{fig:normalizing_EOD2}
\end{figure}

To deduce the longitudinal bunch profile from the measured laser spectrum that is modulated by the Coulomb field of the electron bunch, the modulated spectrum $I_\mathrm{m}$ has to be normalized to the unmodulated laser spectrum $I_\mathrm{u}$ and detector background $I_\mathrm{bg}$ has to be subtracted to get the normalized signal modulation
\begin{equation} 
S_\mathrm{mod} = \frac{I_\mathrm{m}-I_\mathrm{bg}}{I_\mathrm{u}-I_\mathrm{bg}}-1 \quad. 
\label{Smod}
\end{equation} 

The MHz line detector KALYPSO offers the possibility of measuring the modulated and unmodulated laser spectra as well as detector background for every bunch train. 
As an advantage, repetitive noise originating from the laser amplitude can be removed by filtering in Fourier space. 

Figure~\ref{fig:normalizing_EOD1} shows the recorded data for a single bunch train at a repetition rate of \SI{1.13}{MHz} and a bunch charge of \SI{480}{pC}. 
The recording starts 100 frames before the first electron bunch in order to acquire unmodulated laser spectra, which are followed by 500 laser spectra, of which 120 are modulated by electron bunches, and a number of frames without laser pulses present to measure the detector background including some amplified spontaneous emission (ASE) from the laser amplifier. 
The averaged background and averaged unmodulated laser spectrum (Fig.~\ref{fig:normalizing_EOD1}, right) are used to calculate the normalized signal modulation $S_\mathrm{mod}$ (according to Eq.~\ref{Smod}) which is plotted as a color-code image in Fig.~\ref{fig:normalizing_EOD2}~(left) for a bunch train of 120 bunches. 
The half-wave plate in front of the polarizer was set to $\theta=15^\circ$ relative to cross polarization, which is sufficient to have $\Gamma\ll\theta$ and therefor $\Gamma \propto E_\mathrm{THz}$. 
The normalized signal modulation, which is proportional to $E_\mathrm{THz}$ and hence the longitudinal bunch profiles, of three individual electron bunches are depicted in Fig.~\ref{fig:normalizing_EOD2}~(right) together with Gaussian fits to the profiles. 
The measured bunch lengths (rms) vary between \SI{376}{fs} and \SI{396}{fs}.
The electron bunches are followed by wakefields, which are also seen by the GaP crystal. 
Their transverse components lead to a small negative dip after the signal of the electron bunch, followed by additional small signals for several tens of picoseconds. 
Wakefields are caused by diffraction and reflection of the Coulomb field at diameter changes or other impedance changes inside the electron beamline upstream of the GaP crystal.

\begin{figure}
   \centering
   \includegraphics[width=0.9\columnwidth]{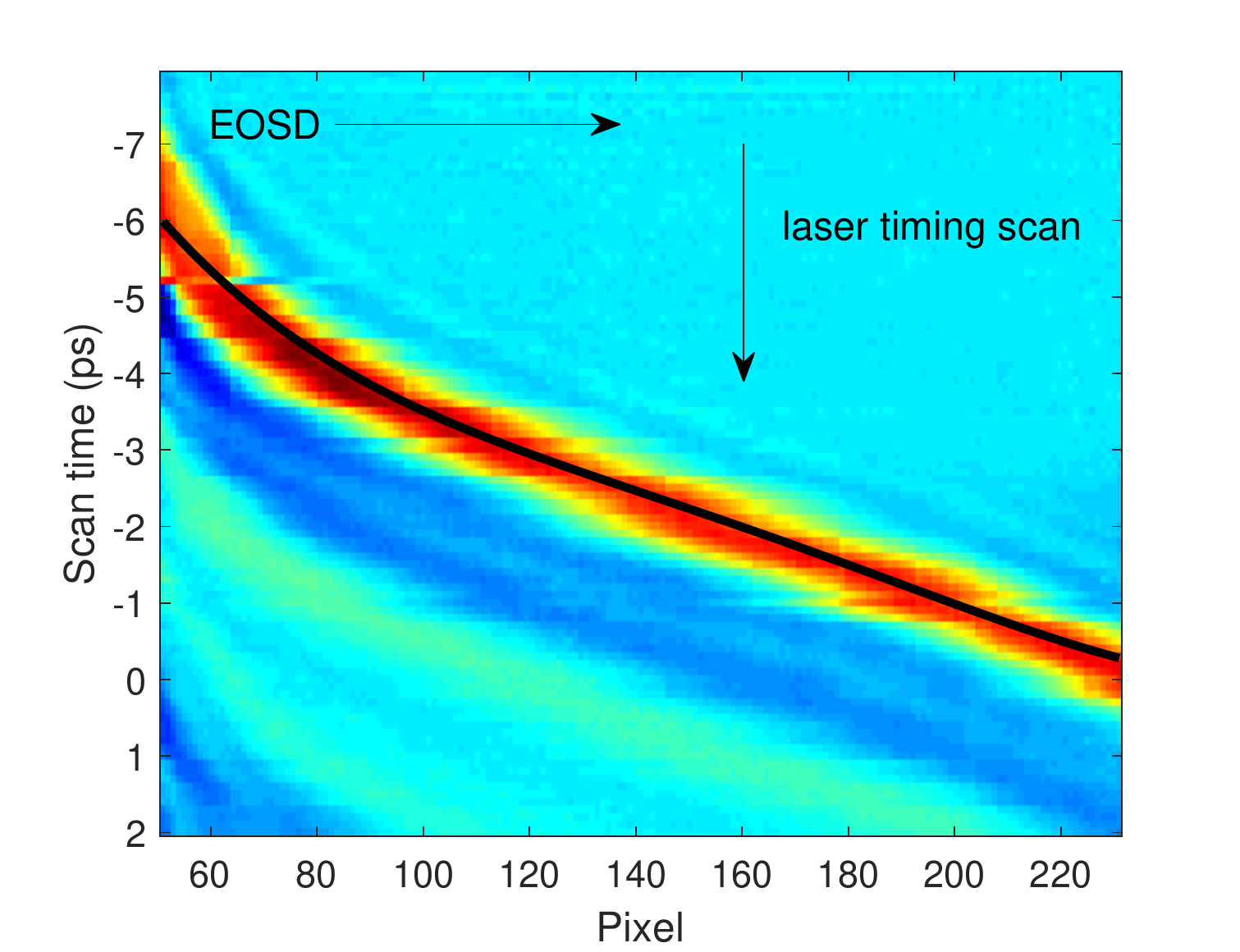}
   \caption{2D plot of single-shot EOD traces (vertical) taken at different laser timings (horizontal). The black line represents the resulting time calibration is obtained from a cubic fit.}
   \label{fig:timecalibration}
\end{figure}

The time calibration applied in Fig.~\ref{fig:normalizing_EOD2} can be deduced from the laser chirp, which is defined by the initial chirp of the pulses from the Yb-fiber laser in combination with the setting of the grating compressor at the accelerator beamline, by scanning the laser pulse relative to the electron bunch at stable accelerator conditions. 
This is demonstrated in Fig.~\ref{fig:timecalibration}.
The laser synchronization allows sub-picosecond time steps with high accuracy, and the resulting shift of the bunch signal in the laser spectrum allows a calibration of each detector pixel to time.
For each EOSD trace, the center position of the main peak is identified and mapped to the corresponding timing from the laser synchronization. 
The resulting time calibration is obtained from a cubic fit and shown as black line in Fig.~\ref{fig:timecalibration}. 

\begin{figure}
  \centering
	\includegraphics[width=0.9\columnwidth]{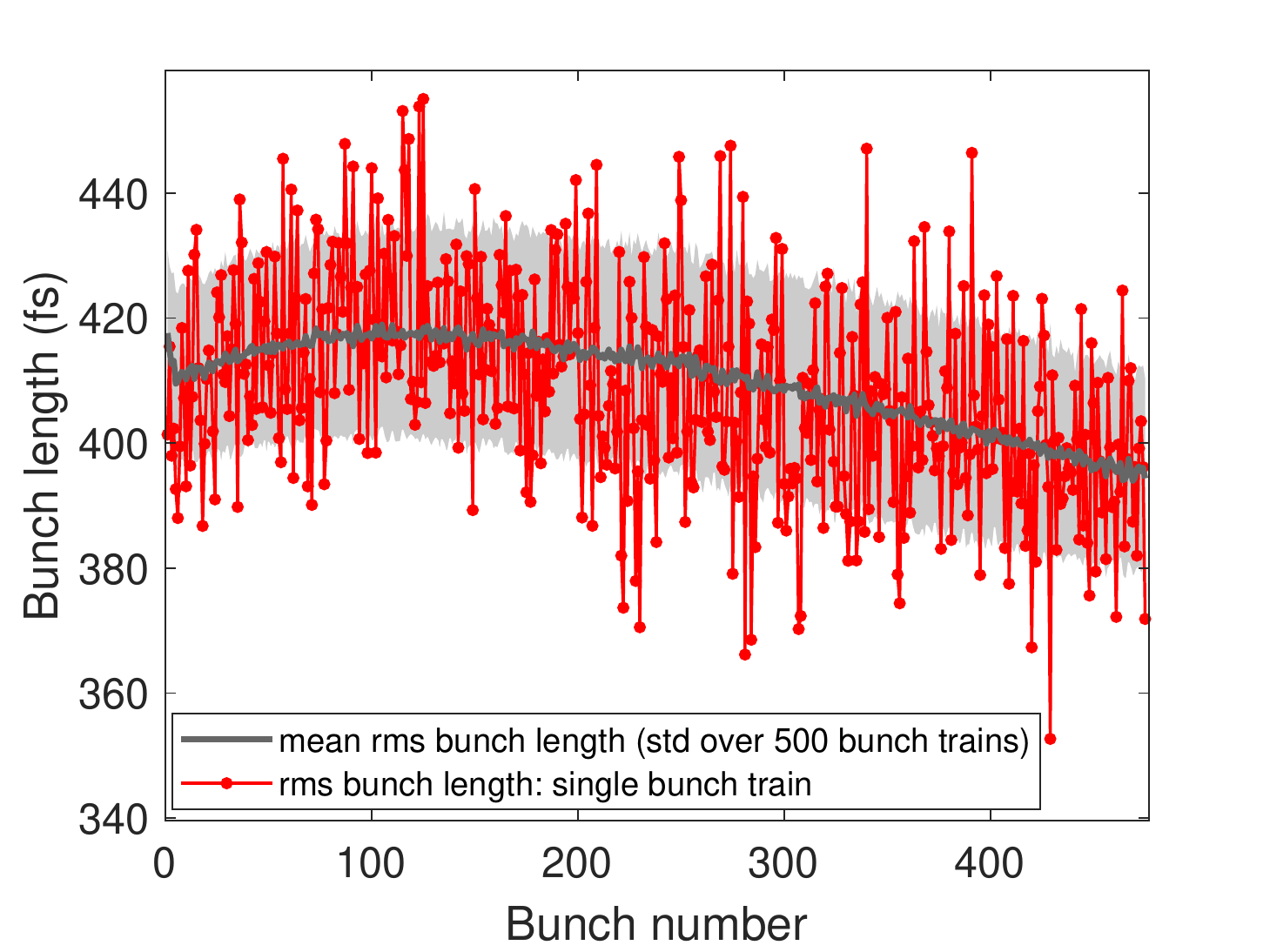}
  	\includegraphics[width=0.9\columnwidth]{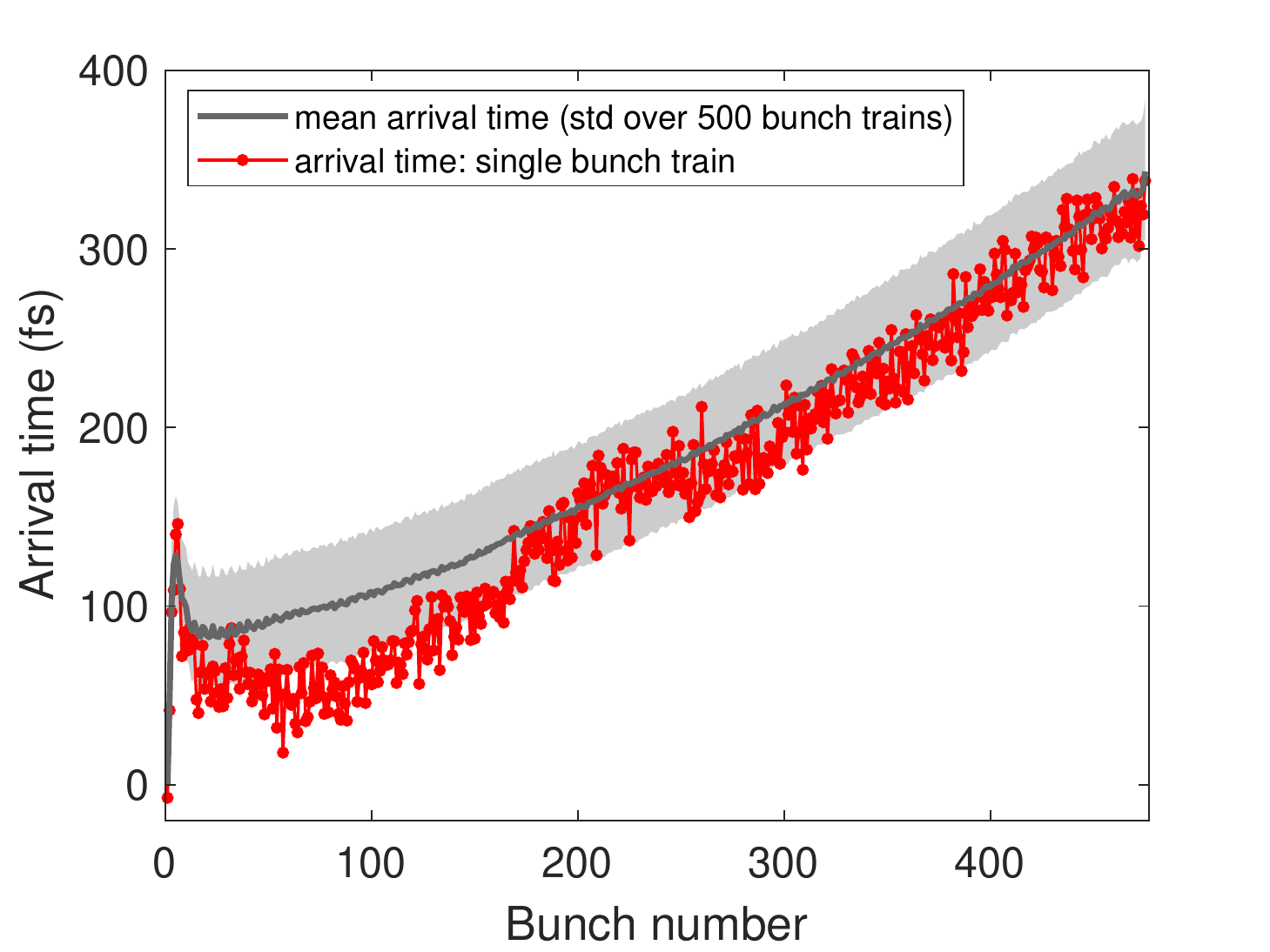}
\caption{Bunch length (rms, top) and arrival time (bottom) of the individual electron bunches (red dots) along the bunch train with a bunch repetition rate of \SI{1.13}{MHz}. The mean values (gray lines) and standard deviation (gray areas) have been calculated from a series of 500 consecutive bunch trains.}
  \label{fig:BLovertrain}
\end{figure}

\subsection{Longitudinal bunch profile and arrival time measurements}
\label{sec:meas-res}

The bunch lengths and arrival times of individual electron bunches in a train of 475 bunches measured at a repetition rate of \SI{1.13}{MHz} and a bunch charge of \SI{240}{pC} are shown in Fig.~\ref{fig:BLovertrain} as red dots together with the mean values (gray lines) and standard deviation (gray areas) of 500 consecutive bunch trains for each bunch number. 
The bunch lengths change from \SI{420}{fs} at bunch number 100 to \SI{400}{fs} towards the end of the bunch train.
These values are larger than the predicted design value about \SI{230}{fs} but in good agreement with predictions from simulation codes for the given parameters of the accelerator. 

The bunch arrival times exhibit a strong shift of about \SI{250}{fs} over the bunch train. 
This is most probably caused by a slope in beam energy over the bunch train, which leads to different beam path lengths in the bunch compressor chicanes. 
The standard deviations of the bunch lengths and bunch arrival times are about \SI{17}{fs} and \SI{35}{fs}, respectively.
 
The measured arrival-time jitter (rms) of about \SI{35}{fs} is composed of contributions from the bunch arrival as well as remaining timing jitter of the laser pulses originating from the laser synchronization to the RF reference.
As the latter contribution has been determined to be about \SI{30}{fs} (see Sec.~\ref{sec:synchronization}) and is thus the main contribution to the measured arrival-time jitter, it can be concluded that the actual bunch arrival-time jitter at this location is considerably smaller.
In addition, any remaining bunch arrival-time jitter originating from the photo-injector will be reduced in the subsequent bunch compression chicane.
Hence, the bunch arrival-time jitter at the undulator beamline, which determines also the arrival-time jitter of the X-ray pulses, is expected to be well below \SI{30}{fs}, which has been corroborated by first results obtained with a bunch-arrival-time monitor that is currently under commissioning.\cite{viti2017}   
First studies\cite{kirkwood2019} on the timing jitter (rms) between the X-ray pulses and a synchronized optical laser for pump-probe experiments yielded a value of $308 \pm 36$~fs, and, in conclusion, the major contribution of this jitter can be attributed to the arrival-time of the pump-probe laser pulses. 
An optical synchronization system\cite{schulz2019} has recently been taken into full operation and will facilitate arrival-time stabilities of below \SI{10}{fs} for the electron bunches and respective X-ray pulses.

\section{Conclusion}
\label{sec:conclusion}

A MHz-repetition-rate EO detection system based on spectral decoding has been developed to detect THz pulses with a single shot resolution of better than \SI{200}{fs}. 
It was designed compact and robust to serve as an electron bunch length and arrival time monitor for full remote operation inside an accelerator environment.
Apart form the optics setup mounted to the electron beamline, all other sub-systems have been designed to fit into \SI{19}{"} chassis for installation in temperature-stabilized and radiation-shielded racks below the electron beamline.

A compact Yb-fiber laser system has been optimized for a broad spectral range which fits the needs of the spectral decoding technique.
Synchronization of the Yb-fiber laser to the RF reference of the accelerator with an accuracy of about \SI{30}{fs} has been demonstrated with a fully-digital controller based on an advanced down-conversion scheme, which has been realized with commercially available MicroTCA.4 components.
The layout of the optics setup, which comprises the vacuum feed-through with the crystal holder inside the accelerator beamline, enables a positioning of the GaP crystal with respect to the electron beam without changing the time delay between the laser pulses and electron bunches. 
The spectrometer has been equipped with the novel MHz line detector KALYPSO to be capable of recording single-shot measurements at repetition rates of up to \SI{2.26}{MHz} over full bunch trains. 
Diagnostics with MHz read-out rates will also become essential for future high repetition rate XFELs.\cite{LCLS2, shine}

First measurement results of the longitudinal bunch profile and arrival time have been obtained downstream of the second bunch compressor chicane at a beam energy of \SI{700}{MeV} and bunch charge of \SI{240}{pC}. 
The measured single-shot bunch lengths range from \SI{400}{fs} to \SI{420}{fs} over the bunch train and are larger than design values but in good agreement with predictions from simulation codes for the actual accelerator settings. 
The measured bunch arrival-time jitter (rms) of about \SI{35}{fs} can mainly be attributed to the timing jitter originating from the laser synchronization and represents an upper limit for the true bunch arrival-time jitter after the second bunch compressor. 

\section{Acknowledgments}

The KALYPSO line detector system has been developed in a collaborative effort between the Karlsruhe Institute of Technology (KIT, Germany), Deutsches Elektronen-Synchrotron (DESY, Germany) and Department of Microelectronics and Computer Science (Lodz University of Technology, Poland). 
The development of the KALYPSO detector has been partially funded by the German Federal Ministry of Education and Research (BMBF) under contract number 05K16VKA.

The authors would like to thank the entire EuXFEL operation team that made these measurements possible.

BS and ChG have designed the entire EOSD setup, carried out the installation and commissioning, and recorded the first measurements.
MF, TK, UM, PP, and KP contributed to the laser to RF synchronization system, and PP optimized the layout of the Yb-fiber laser system.
MC and LR developed the KALYPSO detector board and DRM and AM the corresponding readout electronics for integration into the EuXFEL control system.

\bibliography{RSI_EOD}

\begin{thebibliography}{41}%
\makeatletter
\providecommand \@ifxundefined [1]{%
 \@ifx{#1\undefined}
}%
\providecommand \@ifnum [1]{%
 \ifnum #1\expandafter \@firstoftwo
 \else \expandafter \@secondoftwo
 \fi
}%
\providecommand \@ifx [1]{%
 \ifx #1\expandafter \@firstoftwo
 \else \expandafter \@secondoftwo
 \fi
}%
\providecommand \natexlab [1]{#1}%
\providecommand \enquote  [1]{``#1''}%
\providecommand \bibnamefont  [1]{#1}%
\providecommand \bibfnamefont [1]{#1}%
\providecommand \citenamefont [1]{#1}%
\providecommand \href@noop [0]{\@secondoftwo}%
\providecommand \href [0]{\begingroup \@sanitize@url \@href}%
\providecommand \@href[1]{\@@startlink{#1}\@@href}%
\providecommand \@@href[1]{\endgroup#1\@@endlink}%
\providecommand \@sanitize@url [0]{\catcode `\\12\catcode `\$12\catcode
  `\&12\catcode `\#12\catcode `\^12\catcode `\_12\catcode `\%12\relax}%
\providecommand \@@startlink[1]{}%
\providecommand \@@endlink[0]{}%
\providecommand \url  [0]{\begingroup\@sanitize@url \@url }%
\providecommand \@url [1]{\endgroup\@href {#1}{\urlprefix }}%
\providecommand \urlprefix  [0]{URL }%
\providecommand \Eprint [0]{\href }%
\providecommand \doibase [0]{http://dx.doi.org/}%
\providecommand \selectlanguage [0]{\@gobble}%
\providecommand \bibinfo  [0]{\@secondoftwo}%
\providecommand \bibfield  [0]{\@secondoftwo}%
\providecommand \translation [1]{[#1]}%
\providecommand \BibitemOpen [0]{}%
\providecommand \bibitemStop [0]{}%
\providecommand \bibitemNoStop [0]{.\EOS\space}%
\providecommand \EOS [0]{\spacefactor3000\relax}%
\providecommand \BibitemShut  [1]{\csname bibitem#1\endcsname}%
\let\auto@bib@innerbib\@empty
\bibitem [{\citenamefont {Wu}\ and\ \citenamefont {Zhang}(1995)}]{Wu95}%
  \BibitemOpen
  \bibfield  {author} {\bibinfo {author} {\bibfnamefont {Q.}~\bibnamefont
  {Wu}}\ and\ \bibinfo {author} {\bibfnamefont {X.-C.}\ \bibnamefont {Zhang}},\
  }\bibfield  {title} {\enquote {\bibinfo {title} {Free-space electro-optic
  sampling of terahertz beams},}\ }\href@noop {} {\bibfield  {journal}
  {\bibinfo  {journal} {Appl. Phys. Lett.}\ }\textbf {\bibinfo {volume} {67}},\
  \bibinfo {pages} {3523} (\bibinfo {year} {1995})}\BibitemShut {NoStop}%
\bibitem [{\citenamefont {Shan}\ \emph {et~al.}(2000)\citenamefont {Shan},
  \citenamefont {Weling}, \citenamefont {Knoesel}, \citenamefont {Bartels},
  \citenamefont {Bonn}, \citenamefont {Nahata}, \citenamefont {Reider},\ and\
  \citenamefont {Heinz}}]{shan00}%
  \BibitemOpen
  \bibfield  {author} {\bibinfo {author} {\bibfnamefont {J.}~\bibnamefont
  {Shan}}, \bibinfo {author} {\bibfnamefont {A.~S.}\ \bibnamefont {Weling}},
  \bibinfo {author} {\bibfnamefont {E.}~\bibnamefont {Knoesel}}, \bibinfo
  {author} {\bibfnamefont {L.}~\bibnamefont {Bartels}}, \bibinfo {author}
  {\bibfnamefont {M.}~\bibnamefont {Bonn}}, \bibinfo {author} {\bibfnamefont
  {A.}~\bibnamefont {Nahata}}, \bibinfo {author} {\bibfnamefont {G.~A.}\
  \bibnamefont {Reider}}, \ and\ \bibinfo {author} {\bibfnamefont {T.~F.}\
  \bibnamefont {Heinz}},\ }\bibfield  {title} {\enquote {\bibinfo {title}
  {Single-shot measurements of terahertz electromagnetic pulses by the use of
  electro-optic sampling},}\ }\href@noop {} {\bibfield  {journal} {\bibinfo
  {journal} {Optics Letters}\ }\textbf {\bibinfo {volume} {25}},\ \bibinfo
  {pages} {436--428} (\bibinfo {year} {2000})}\BibitemShut {NoStop}%
\bibitem [{\citenamefont {Jiang}\ and\ \citenamefont {Zhang}(1998)}]{jiang98}%
  \BibitemOpen
  \bibfield  {author} {\bibinfo {author} {\bibfnamefont {Z.}~\bibnamefont
  {Jiang}}\ and\ \bibinfo {author} {\bibfnamefont {X.-C.}\ \bibnamefont
  {Zhang}},\ }\bibfield  {title} {\enquote {\bibinfo {title} {Electro-optic
  measurement of {THz} field pulses with a chirped optical beam},}\ }\href@noop
  {} {\bibfield  {journal} {\bibinfo  {journal} {Appl. Phys. Lett.}\ }\textbf
  {\bibinfo {volume} {72}},\ \bibinfo {pages} {1945--1947} (\bibinfo {year}
  {1998})}\BibitemShut {NoStop}%
\bibitem [{\citenamefont {Jamison}\ \emph {et~al.}(2003)\citenamefont
  {Jamison}, \citenamefont {Shen}, \citenamefont {MacLeod}, \citenamefont
  {Gillespie},\ and\ \citenamefont {Jaroszynski}}]{Jamison03}%
  \BibitemOpen
  \bibfield  {author} {\bibinfo {author} {\bibfnamefont {S.~P.}\ \bibnamefont
  {Jamison}}, \bibinfo {author} {\bibfnamefont {J.}~\bibnamefont {Shen}},
  \bibinfo {author} {\bibfnamefont {A.~M.}\ \bibnamefont {MacLeod}}, \bibinfo
  {author} {\bibfnamefont {W.~A.}\ \bibnamefont {Gillespie}}, \ and\ \bibinfo
  {author} {\bibfnamefont {D.~A.}\ \bibnamefont {Jaroszynski}},\ }\bibfield
  {title} {\enquote {\bibinfo {title} {High-temporal-resolution, single-shot
  characterization of terahertz pulses},}\ }\href@noop {} {\bibfield  {journal}
  {\bibinfo  {journal} {Optics Letters}\ }\textbf {\bibinfo {volume} {28}},\
  \bibinfo {pages} {1710--1712} (\bibinfo {year} {2003})}\BibitemShut {NoStop}%
\bibitem [{\citenamefont {Yasumatsu}\ and\ \citenamefont
  {Watanabe}(2012)}]{Yasumatsu2012}%
  \BibitemOpen
  \bibfield  {author} {\bibinfo {author} {\bibfnamefont {N.}~\bibnamefont
  {Yasumatsu}}\ and\ \bibinfo {author} {\bibfnamefont {S.}~\bibnamefont
  {Watanabe}},\ }\bibfield  {title} {\enquote {\bibinfo {title} {Precise
  real-time polarization measurement of terahertz electromagnetic waves by a
  spinning electro-optic sensor},}\ }\href {\doibase 10.1063/1.3683570}
  {\bibfield  {journal} {\bibinfo  {journal} {Review of Scientific
  Instruments}\ }\textbf {\bibinfo {volume} {83}},\ \bibinfo {pages} {023104}
  (\bibinfo {year} {2012})},\ \Eprint
  {http://arxiv.org/abs/https://doi.org/10.1063/1.3683570}
  {https://doi.org/10.1063/1.3683570} \BibitemShut {NoStop}%
\bibitem [{\citenamefont {Riek}\ \emph {et~al.}(2015)\citenamefont {Riek},
  \citenamefont {Seletskiy}, \citenamefont {Moskalenko}, \citenamefont
  {Schmidt}, \citenamefont {Krauspe}, \citenamefont {Eckart}, \citenamefont
  {Eggert}, \citenamefont {Burkard},\ and\ \citenamefont
  {Leitenstorfer}}]{Riek2015}%
  \BibitemOpen
  \bibfield  {author} {\bibinfo {author} {\bibfnamefont {C.}~\bibnamefont
  {Riek}}, \bibinfo {author} {\bibfnamefont {D.~V.}\ \bibnamefont {Seletskiy}},
  \bibinfo {author} {\bibfnamefont {A.~S.}\ \bibnamefont {Moskalenko}},
  \bibinfo {author} {\bibfnamefont {J.~F.}\ \bibnamefont {Schmidt}}, \bibinfo
  {author} {\bibfnamefont {P.}~\bibnamefont {Krauspe}}, \bibinfo {author}
  {\bibfnamefont {S.}~\bibnamefont {Eckart}}, \bibinfo {author} {\bibfnamefont
  {S.}~\bibnamefont {Eggert}}, \bibinfo {author} {\bibfnamefont
  {G.}~\bibnamefont {Burkard}}, \ and\ \bibinfo {author} {\bibfnamefont
  {A.}~\bibnamefont {Leitenstorfer}},\ }\bibfield  {title} {\enquote {\bibinfo
  {title} {Direct sampling of electric-field vacuum fluctuations},}\ }\href
  {\doibase 10.1126/science.aac9788} {\bibfield  {journal} {\bibinfo  {journal}
  {Science}\ }\textbf {\bibinfo {volume} {350}},\ \bibinfo {pages} {420--423}
  (\bibinfo {year} {2015})},\ \Eprint
  {http://arxiv.org/abs/https://science.sciencemag.org/content/350/6259/420.full.pdf}
  {https://science.sciencemag.org/content/350/6259/420.full.pdf} \BibitemShut
  {NoStop}%
\bibitem [{\citenamefont {Sanjuan}, \citenamefont {Gaborit},\ and\
  \citenamefont {Coutaz}(2018)}]{Sanjuan2018}%
  \BibitemOpen
  \bibfield  {author} {\bibinfo {author} {\bibfnamefont {F.}~\bibnamefont
  {Sanjuan}}, \bibinfo {author} {\bibfnamefont {G.}~\bibnamefont {Gaborit}}, \
  and\ \bibinfo {author} {\bibfnamefont {J.-L.}\ \bibnamefont {Coutaz}},\
  }\bibfield  {title} {\enquote {\bibinfo {title} {Full electro-optic terahertz
  time-domain spectrometer for polarimetric studies},}\ }\href {\doibase
  10.1364/AO.57.006055} {\bibfield  {journal} {\bibinfo  {journal} {Appl.
  Opt.}\ }\textbf {\bibinfo {volume} {57}},\ \bibinfo {pages} {6055--6060}
  (\bibinfo {year} {2018})}\BibitemShut {NoStop}%
\bibitem [{\citenamefont {Yan}\ \emph {et~al.}(2000)\citenamefont {Yan},
  \citenamefont {MacLeod}, \citenamefont {Gillespie}, \citenamefont {Knippels},
  \citenamefont {Oepts}, \citenamefont {van~der Meer},\ and\ \citenamefont
  {Seidel}}]{yan00}%
  \BibitemOpen
  \bibfield  {author} {\bibinfo {author} {\bibfnamefont {X.}~\bibnamefont
  {Yan}}, \bibinfo {author} {\bibfnamefont {A.~M.}\ \bibnamefont {MacLeod}},
  \bibinfo {author} {\bibfnamefont {W.~A.}\ \bibnamefont {Gillespie}}, \bibinfo
  {author} {\bibfnamefont {G.~M.~H.}\ \bibnamefont {Knippels}}, \bibinfo
  {author} {\bibfnamefont {D.}~\bibnamefont {Oepts}}, \bibinfo {author}
  {\bibfnamefont {A.~F.~G.}\ \bibnamefont {van~der Meer}}, \ and\ \bibinfo
  {author} {\bibfnamefont {W.}~\bibnamefont {Seidel}},\ }\bibfield  {title}
  {\enquote {\bibinfo {title} {Subpicosecond electro-optic measurement of
  relativistic electron pulses},}\ }\href {\doibase
  10.1103/PhysRevLett.85.3404} {\bibfield  {journal} {\bibinfo  {journal}
  {Phys. Rev. Lett.}\ }\textbf {\bibinfo {volume} {85}},\ \bibinfo {pages}
  {3404--3407} (\bibinfo {year} {2000})}\BibitemShut {NoStop}%
\bibitem [{\citenamefont {Wilke}\ \emph {et~al.}(2002)\citenamefont {Wilke},
  \citenamefont {MacLeod}, \citenamefont {Gillespie}, \citenamefont {Berden},
  \citenamefont {Knippels},\ and\ \citenamefont {van~der Meer}}]{wilke02}%
  \BibitemOpen
  \bibfield  {author} {\bibinfo {author} {\bibfnamefont {I.}~\bibnamefont
  {Wilke}}, \bibinfo {author} {\bibfnamefont {A.~M.}\ \bibnamefont {MacLeod}},
  \bibinfo {author} {\bibfnamefont {W.~A.}\ \bibnamefont {Gillespie}}, \bibinfo
  {author} {\bibfnamefont {G.}~\bibnamefont {Berden}}, \bibinfo {author}
  {\bibfnamefont {G.~M.~H.}\ \bibnamefont {Knippels}}, \ and\ \bibinfo {author}
  {\bibfnamefont {A.~F.~G.}\ \bibnamefont {van~der Meer}},\ }\bibfield  {title}
  {\enquote {\bibinfo {title} {Single-shot electron-beam bunch length
  measurements},}\ }\href {\doibase 10.1103/PhysRevLett.88.124801} {\bibfield
  {journal} {\bibinfo  {journal} {Phys. Rev. Lett.}\ }\textbf {\bibinfo
  {volume} {88}},\ \bibinfo {pages} {124801} (\bibinfo {year}
  {2002})}\BibitemShut {NoStop}%
\bibitem [{\citenamefont {Berden}\ \emph {et~al.}(2004)\citenamefont {Berden},
  \citenamefont {Jamison}, \citenamefont {McLeod}, \citenamefont {Gillespie},
  \citenamefont {Redlich},\ and\ \citenamefont {van~der Meer}}]{berden04}%
  \BibitemOpen
  \bibfield  {author} {\bibinfo {author} {\bibfnamefont {G.}~\bibnamefont
  {Berden}}, \bibinfo {author} {\bibfnamefont {S.~P.}\ \bibnamefont {Jamison}},
  \bibinfo {author} {\bibfnamefont {A.~M.}\ \bibnamefont {McLeod}}, \bibinfo
  {author} {\bibfnamefont {W.~A.}\ \bibnamefont {Gillespie}}, \bibinfo {author}
  {\bibfnamefont {B.}~\bibnamefont {Redlich}}, \ and\ \bibinfo {author}
  {\bibfnamefont {A.~F.~G.}\ \bibnamefont {van~der Meer}},\ }\bibfield  {title}
  {\enquote {\bibinfo {title} {Electro-optic technique with improved time
  resolution for real-time, nondestructive, single-shot measurements of
  femtosecond electron bunch profiles},}\ }\href@noop {} {\bibfield  {journal}
  {\bibinfo  {journal} {Phys. Rev. Lett.}\ }\textbf {\bibinfo {volume} {93}},\
  \bibinfo {pages} {114802} (\bibinfo {year} {2004})}\BibitemShut {NoStop}%
\bibitem [{\citenamefont {Cavalieri}\ \emph {et~al.}(2005)\citenamefont
  {Cavalieri} \emph {et~al.}}]{cavalieri05}%
  \BibitemOpen
  \bibfield  {author} {\bibinfo {author} {\bibfnamefont {A.~L.}\ \bibnamefont
  {Cavalieri}} \emph {et~al.},\ }\bibfield  {title} {\enquote {\bibinfo {title}
  {Clocking femtosecond x rays},}\ }\href
  {http://link.aps.org/abstract/PRL/v94/e114801} {\bibfield  {journal}
  {\bibinfo  {journal} {Phys. Rev. Lett.}\ }\textbf {\bibinfo {volume} {94}},\
  \bibinfo {eid} {114801} (\bibinfo {year} {2005})}\BibitemShut {NoStop}%
\bibitem [{\citenamefont {Steffen}\ \emph {et~al.}(2009)\citenamefont
  {Steffen}, \citenamefont {Arsov}, \citenamefont {Berden}, \citenamefont
  {Gillespie}, \citenamefont {Jamison}, \citenamefont {MacLeod}, \citenamefont
  {Van Der~Meer}, \citenamefont {Phillips}, \citenamefont {Schlarb},
  \citenamefont {Schmidt},\ and\ \citenamefont {Schm\"user}}]{steffen09a}%
  \BibitemOpen
  \bibfield  {author} {\bibinfo {author} {\bibfnamefont {B.}~\bibnamefont
  {Steffen}}, \bibinfo {author} {\bibfnamefont {V.}~\bibnamefont {Arsov}},
  \bibinfo {author} {\bibfnamefont {G.}~\bibnamefont {Berden}}, \bibinfo
  {author} {\bibfnamefont {W.~A.}\ \bibnamefont {Gillespie}}, \bibinfo {author}
  {\bibfnamefont {S.~P.}\ \bibnamefont {Jamison}}, \bibinfo {author}
  {\bibfnamefont {A.~M.}\ \bibnamefont {MacLeod}}, \bibinfo {author}
  {\bibfnamefont {A.~F.~G.}\ \bibnamefont {Van Der~Meer}}, \bibinfo {author}
  {\bibfnamefont {P.~J.}\ \bibnamefont {Phillips}}, \bibinfo {author}
  {\bibfnamefont {H.}~\bibnamefont {Schlarb}}, \bibinfo {author} {\bibfnamefont
  {B.}~\bibnamefont {Schmidt}}, \ and\ \bibinfo {author} {\bibfnamefont
  {P.}~\bibnamefont {Schm\"user}},\ }\bibfield  {title} {{\selectlanguage
  {English}\enquote {\bibinfo {title} {Electro-optic time profile monitors for
  femtosecond electron bunches at the soft x-ray free-electron laser
  {FLASH}},}\ }}\href@noop {} {\bibfield  {journal} {\bibinfo  {journal}
  {Physical Review Special Topics - Accelerators and Beams}\ }\textbf {\bibinfo
  {volume} {12}} (\bibinfo {year} {2009})}\BibitemShut {NoStop}%
\bibitem [{\citenamefont {Funkner}\ \emph {et~al.}(2019)\citenamefont
  {Funkner}, \citenamefont {Blomley}, \citenamefont {Br\"undermann},
  \citenamefont {Caselle}, \citenamefont {Hiller}, \citenamefont {Nasse},
  \citenamefont {Niehues}, \citenamefont {Rota}, \citenamefont {Sch\"onfeldt},
  \citenamefont {Walther}, \citenamefont {Weber},\ and\ \citenamefont
  {M\"uller}}]{funkner2019}%
  \BibitemOpen
  \bibfield  {author} {\bibinfo {author} {\bibfnamefont {S.}~\bibnamefont
  {Funkner}}, \bibinfo {author} {\bibfnamefont {E.}~\bibnamefont {Blomley}},
  \bibinfo {author} {\bibfnamefont {E.}~\bibnamefont {Br\"undermann}}, \bibinfo
  {author} {\bibfnamefont {M.}~\bibnamefont {Caselle}}, \bibinfo {author}
  {\bibfnamefont {N.}~\bibnamefont {Hiller}}, \bibinfo {author} {\bibfnamefont
  {M.~J.}\ \bibnamefont {Nasse}}, \bibinfo {author} {\bibfnamefont
  {G.}~\bibnamefont {Niehues}}, \bibinfo {author} {\bibfnamefont
  {L.}~\bibnamefont {Rota}}, \bibinfo {author} {\bibfnamefont {P.}~\bibnamefont
  {Sch\"onfeldt}}, \bibinfo {author} {\bibfnamefont {S.}~\bibnamefont
  {Walther}}, \bibinfo {author} {\bibfnamefont {M.}~\bibnamefont {Weber}}, \
  and\ \bibinfo {author} {\bibfnamefont {A.-S.}\ \bibnamefont {M\"uller}},\
  }\bibfield  {title} {\enquote {\bibinfo {title} {High throughput data
  streaming of individual longitudinal electron bunch profiles},}\ }\href
  {\doibase 10.1103/PhysRevAccelBeams.22.022801} {\bibfield  {journal}
  {\bibinfo  {journal} {Phys. Rev. Accel. Beams}\ }\textbf {\bibinfo {volume}
  {22}},\ \bibinfo {pages} {022801} (\bibinfo {year} {2019})}\BibitemShut
  {NoStop}%
\bibitem [{\citenamefont {Altarelli}(2015)}]{altarelli2015}%
  \BibitemOpen
  \bibfield  {author} {\bibinfo {author} {\bibfnamefont {M.}~\bibnamefont
  {Altarelli}},\ }\bibfield  {title} {\enquote {\bibinfo {title} {The {European
  X-ray Free-Electron Laser}: toward an ultra-bright, high repetition-rate
  x-ray source},}\ }\href {\doibase 10.1017/hpl.2015.17} {\bibfield  {journal}
  {\bibinfo  {journal} {High Power Laser Science and Engineering}\ }\textbf
  {\bibinfo {volume} {3}},\ \bibinfo {pages} {e18} (\bibinfo {year}
  {2015})}\BibitemShut {NoStop}%
\bibitem [{\citenamefont {Vagovič}\ \emph {et~al.}(2019)\citenamefont
  {Vagovič}, \citenamefont {Sato}, \citenamefont {Mikeš}, \citenamefont
  {Mills}, \citenamefont {Graceffa}, \citenamefont {Mattsson}, \citenamefont
  {Villanueva-Perez}, \citenamefont {Ershov}, \citenamefont {Faragó},
  \citenamefont {Uličný}, \citenamefont {Kirkwood}, \citenamefont {Letrun},
  \citenamefont {Mokso}, \citenamefont {Zdora}, \citenamefont {Olbinado},
  \citenamefont {Rack}, \citenamefont {Baumbach}, \citenamefont {Schulz},
  \citenamefont {Meents}, \citenamefont {Chapman},\ and\ \citenamefont
  {Mancuso}}]{vagovic2019}%
  \BibitemOpen
  \bibfield  {author} {\bibinfo {author} {\bibfnamefont {P.}~\bibnamefont
  {Vagovič}}, \bibinfo {author} {\bibfnamefont {T.}~\bibnamefont {Sato}},
  \bibinfo {author} {\bibfnamefont {L.}~\bibnamefont {Mikeš}}, \bibinfo
  {author} {\bibfnamefont {G.}~\bibnamefont {Mills}}, \bibinfo {author}
  {\bibfnamefont {R.}~\bibnamefont {Graceffa}}, \bibinfo {author}
  {\bibfnamefont {F.}~\bibnamefont {Mattsson}}, \bibinfo {author}
  {\bibfnamefont {P.}~\bibnamefont {Villanueva-Perez}}, \bibinfo {author}
  {\bibfnamefont {A.}~\bibnamefont {Ershov}}, \bibinfo {author} {\bibfnamefont
  {T.}~\bibnamefont {Faragó}}, \bibinfo {author} {\bibfnamefont
  {J.}~\bibnamefont {Uličný}}, \bibinfo {author} {\bibfnamefont
  {H.}~\bibnamefont {Kirkwood}}, \bibinfo {author} {\bibfnamefont
  {R.}~\bibnamefont {Letrun}}, \bibinfo {author} {\bibfnamefont
  {R.}~\bibnamefont {Mokso}}, \bibinfo {author} {\bibfnamefont {M.-C.}\
  \bibnamefont {Zdora}}, \bibinfo {author} {\bibfnamefont {M.~P.}\ \bibnamefont
  {Olbinado}}, \bibinfo {author} {\bibfnamefont {A.}~\bibnamefont {Rack}},
  \bibinfo {author} {\bibfnamefont {T.}~\bibnamefont {Baumbach}}, \bibinfo
  {author} {\bibfnamefont {J.}~\bibnamefont {Schulz}}, \bibinfo {author}
  {\bibfnamefont {A.}~\bibnamefont {Meents}}, \bibinfo {author} {\bibfnamefont
  {H.~N.}\ \bibnamefont {Chapman}}, \ and\ \bibinfo {author} {\bibfnamefont
  {A.~P.}\ \bibnamefont {Mancuso}},\ }\bibfield  {title} {\enquote {\bibinfo
  {title} {Megahertz x-ray microscopy at x-ray free-electron laser and
  synchrotron sources},}\ }\href {http://xfel.tind.io/record/1974} {\bibfield
  {journal} {\bibinfo  {journal} {Optica}\ } (\bibinfo {year}
  {2019})}\BibitemShut {NoStop}%
\bibitem [{\citenamefont {Gisriel}\ \emph {et~al.}(2019)\citenamefont
  {Gisriel}, \citenamefont {Coe}, \citenamefont {Letrun}, \citenamefont
  {Yefanov}, \citenamefont {Luna-Chavez}, \citenamefont {Stander},
  \citenamefont {Lisova}, \citenamefont {Mariani}, \citenamefont {Kuhn},
  \citenamefont {Aplin}, \citenamefont {Grant}, \citenamefont {Dörner},
  \citenamefont {Sato}, \citenamefont {Echelmeier}, \citenamefont
  {Cruz~Villarreal}, \citenamefont {Hunter}, \citenamefont {Wiedorn},
  \citenamefont {Knoska}, \citenamefont {Mazalova}, \citenamefont
  {Roy-Chowdhury}, \citenamefont {Yang}, \citenamefont {Jones}, \citenamefont
  {Bean}, \citenamefont {Bielecki}, \citenamefont {Kim}, \citenamefont {Mills},
  \citenamefont {Weinhausen}, \citenamefont {Meza}, \citenamefont {Al-Qudami},
  \citenamefont {Bajt}, \citenamefont {Brehm}, \citenamefont {Botha},
  \citenamefont {Boukhelef}, \citenamefont {Brockhauser}, \citenamefont
  {Bruce}, \citenamefont {Coleman}, \citenamefont {Danilevski}, \citenamefont
  {Discianno}, \citenamefont {Dobson}, \citenamefont {Fangohr}, \citenamefont
  {Martin-Garcia}, \citenamefont {Gevorkov}, \citenamefont {Hauf},
  \citenamefont {Hosseinizadeh}, \citenamefont {Januschek}, \citenamefont
  {Ketawala}, \citenamefont {Kupitz}, \citenamefont {Maia}, \citenamefont
  {Manetti}, \citenamefont {Messerschmidt}, \citenamefont {Michelat},
  \citenamefont {Mondal}, \citenamefont {Ourmazd}, \citenamefont {Previtali},
  \citenamefont {Sarrou}, \citenamefont {Schön}, \citenamefont {Schwander},
  \citenamefont {Shelby}, \citenamefont {Silenzi}, \citenamefont
  {Sztuk-Dambietz}, \citenamefont {Szuba}, \citenamefont {Turcato},
  \citenamefont {White}, \citenamefont {Wrona}, \citenamefont {Xu},
  \citenamefont {Abdellatif}, \citenamefont {Zook}, \citenamefont {Spence},
  \citenamefont {Chapman}, \citenamefont {Barty}, \citenamefont {Kirian},
  \citenamefont {Frank}, \citenamefont {Ros}, \citenamefont {Schmidt},
  \citenamefont {Fromme}, \citenamefont {Mancuso}, \citenamefont {Fromme},\
  and\ \citenamefont {Zatsepin}}]{gisriel2019}%
  \BibitemOpen
  \bibfield  {author} {\bibinfo {author} {\bibfnamefont {C.}~\bibnamefont
  {Gisriel}}, \bibinfo {author} {\bibfnamefont {J.}~\bibnamefont {Coe}},
  \bibinfo {author} {\bibfnamefont {R.}~\bibnamefont {Letrun}}, \bibinfo
  {author} {\bibfnamefont {O.~M.}\ \bibnamefont {Yefanov}}, \bibinfo {author}
  {\bibfnamefont {C.}~\bibnamefont {Luna-Chavez}}, \bibinfo {author}
  {\bibfnamefont {N.~E.}\ \bibnamefont {Stander}}, \bibinfo {author}
  {\bibfnamefont {S.}~\bibnamefont {Lisova}}, \bibinfo {author} {\bibfnamefont
  {V.}~\bibnamefont {Mariani}}, \bibinfo {author} {\bibfnamefont
  {M.}~\bibnamefont {Kuhn}}, \bibinfo {author} {\bibfnamefont {S.}~\bibnamefont
  {Aplin}}, \bibinfo {author} {\bibfnamefont {T.~D.}\ \bibnamefont {Grant}},
  \bibinfo {author} {\bibfnamefont {K.}~\bibnamefont {Dörner}}, \bibinfo
  {author} {\bibfnamefont {T.}~\bibnamefont {Sato}}, \bibinfo {author}
  {\bibfnamefont {A.}~\bibnamefont {Echelmeier}}, \bibinfo {author}
  {\bibfnamefont {J.}~\bibnamefont {Cruz~Villarreal}}, \bibinfo {author}
  {\bibfnamefont {M.~S.}\ \bibnamefont {Hunter}}, \bibinfo {author}
  {\bibfnamefont {M.~O.}\ \bibnamefont {Wiedorn}}, \bibinfo {author}
  {\bibfnamefont {J.}~\bibnamefont {Knoska}}, \bibinfo {author} {\bibfnamefont
  {V.}~\bibnamefont {Mazalova}}, \bibinfo {author} {\bibfnamefont
  {S.}~\bibnamefont {Roy-Chowdhury}}, \bibinfo {author} {\bibfnamefont {J.-H.}\
  \bibnamefont {Yang}}, \bibinfo {author} {\bibfnamefont {A.}~\bibnamefont
  {Jones}}, \bibinfo {author} {\bibfnamefont {R.}~\bibnamefont {Bean}},
  \bibinfo {author} {\bibfnamefont {J.}~\bibnamefont {Bielecki}}, \bibinfo
  {author} {\bibfnamefont {Y.}~\bibnamefont {Kim}}, \bibinfo {author}
  {\bibfnamefont {G.}~\bibnamefont {Mills}}, \bibinfo {author} {\bibfnamefont
  {B.}~\bibnamefont {Weinhausen}}, \bibinfo {author} {\bibfnamefont {J.~D.}\
  \bibnamefont {Meza}}, \bibinfo {author} {\bibfnamefont {N.}~\bibnamefont
  {Al-Qudami}}, \bibinfo {author} {\bibfnamefont {S.}~\bibnamefont {Bajt}},
  \bibinfo {author} {\bibfnamefont {G.}~\bibnamefont {Brehm}}, \bibinfo
  {author} {\bibfnamefont {S.}~\bibnamefont {Botha}}, \bibinfo {author}
  {\bibfnamefont {D.}~\bibnamefont {Boukhelef}}, \bibinfo {author}
  {\bibfnamefont {S.}~\bibnamefont {Brockhauser}}, \bibinfo {author}
  {\bibfnamefont {B.~D.}\ \bibnamefont {Bruce}}, \bibinfo {author}
  {\bibfnamefont {M.~A.}\ \bibnamefont {Coleman}}, \bibinfo {author}
  {\bibfnamefont {C.}~\bibnamefont {Danilevski}}, \bibinfo {author}
  {\bibfnamefont {E.}~\bibnamefont {Discianno}}, \bibinfo {author}
  {\bibfnamefont {Z.}~\bibnamefont {Dobson}}, \bibinfo {author} {\bibfnamefont
  {H.}~\bibnamefont {Fangohr}}, \bibinfo {author} {\bibfnamefont {J.~M.}\
  \bibnamefont {Martin-Garcia}}, \bibinfo {author} {\bibfnamefont
  {Y.}~\bibnamefont {Gevorkov}}, \bibinfo {author} {\bibfnamefont
  {S.}~\bibnamefont {Hauf}}, \bibinfo {author} {\bibfnamefont {A.}~\bibnamefont
  {Hosseinizadeh}}, \bibinfo {author} {\bibfnamefont {F.}~\bibnamefont
  {Januschek}}, \bibinfo {author} {\bibfnamefont {G.~K.}\ \bibnamefont
  {Ketawala}}, \bibinfo {author} {\bibfnamefont {C.}~\bibnamefont {Kupitz}},
  \bibinfo {author} {\bibfnamefont {L.}~\bibnamefont {Maia}}, \bibinfo {author}
  {\bibfnamefont {M.}~\bibnamefont {Manetti}}, \bibinfo {author} {\bibfnamefont
  {M.}~\bibnamefont {Messerschmidt}}, \bibinfo {author} {\bibfnamefont
  {T.}~\bibnamefont {Michelat}}, \bibinfo {author} {\bibfnamefont
  {J.}~\bibnamefont {Mondal}}, \bibinfo {author} {\bibfnamefont
  {A.}~\bibnamefont {Ourmazd}}, \bibinfo {author} {\bibfnamefont
  {G.}~\bibnamefont {Previtali}}, \bibinfo {author} {\bibfnamefont
  {I.}~\bibnamefont {Sarrou}}, \bibinfo {author} {\bibfnamefont
  {S.}~\bibnamefont {Schön}}, \bibinfo {author} {\bibfnamefont
  {P.}~\bibnamefont {Schwander}}, \bibinfo {author} {\bibfnamefont {M.~L.}\
  \bibnamefont {Shelby}}, \bibinfo {author} {\bibfnamefont {A.}~\bibnamefont
  {Silenzi}}, \bibinfo {author} {\bibfnamefont {J.}~\bibnamefont
  {Sztuk-Dambietz}}, \bibinfo {author} {\bibfnamefont {J.}~\bibnamefont
  {Szuba}}, \bibinfo {author} {\bibfnamefont {M.}~\bibnamefont {Turcato}},
  \bibinfo {author} {\bibfnamefont {T.~A.}\ \bibnamefont {White}}, \bibinfo
  {author} {\bibfnamefont {K.}~\bibnamefont {Wrona}}, \bibinfo {author}
  {\bibfnamefont {C.}~\bibnamefont {Xu}}, \bibinfo {author} {\bibfnamefont
  {M.~H.}\ \bibnamefont {Abdellatif}}, \bibinfo {author} {\bibfnamefont
  {J.~D.}\ \bibnamefont {Zook}}, \bibinfo {author} {\bibfnamefont {J.~C.~H.}\
  \bibnamefont {Spence}}, \bibinfo {author} {\bibfnamefont {H.~N.}\
  \bibnamefont {Chapman}}, \bibinfo {author} {\bibfnamefont {A.}~\bibnamefont
  {Barty}}, \bibinfo {author} {\bibfnamefont {R.~A.}\ \bibnamefont {Kirian}},
  \bibinfo {author} {\bibfnamefont {M.}~\bibnamefont {Frank}}, \bibinfo
  {author} {\bibfnamefont {A.}~\bibnamefont {Ros}}, \bibinfo {author}
  {\bibfnamefont {M.}~\bibnamefont {Schmidt}}, \bibinfo {author} {\bibfnamefont
  {R.}~\bibnamefont {Fromme}}, \bibinfo {author} {\bibfnamefont {A.~P.}\
  \bibnamefont {Mancuso}}, \bibinfo {author} {\bibfnamefont {P.}~\bibnamefont
  {Fromme}}, \ and\ \bibinfo {author} {\bibfnamefont {N.~A.}\ \bibnamefont
  {Zatsepin}},\ }\bibfield  {title} {\enquote {\bibinfo {title} {Membrane
  protein megahertz crystallography at the {European XFEL}},}\ }\href
  {http://xfel.tind.io/record/1986} {\bibfield  {journal} {\bibinfo  {journal}
  {Nature Communications}\ } (\bibinfo {year} {2019})}\BibitemShut {NoStop}%
\bibitem [{\citenamefont {Pandey}\ \emph {et~al.}(2019)\citenamefont {Pandey},
  \citenamefont {Bean}, \citenamefont {Sato}, \citenamefont {Poudyal},
  \citenamefont {Bielecki}, \citenamefont {Cruz~Villarreal}, \citenamefont
  {Yefanov}, \citenamefont {Mariani}, \citenamefont {White}, \citenamefont
  {Kupitz}, \citenamefont {Hunter}, \citenamefont {Abdellatif}, \citenamefont
  {Bajt}, \citenamefont {Bondar}, \citenamefont {Echelmeier}, \citenamefont
  {Doppler}, \citenamefont {Emons}, \citenamefont {Frank}, \citenamefont
  {Fromme}, \citenamefont {Gevorkov}, \citenamefont {Giovanetti}, \citenamefont
  {Jiang}, \citenamefont {Kim}, \citenamefont {Kim}, \citenamefont {Kirkwood},
  \citenamefont {Klimovskaia}, \citenamefont {Knoska}, \citenamefont {Koua},
  \citenamefont {Letrun}, \citenamefont {Lisova}, \citenamefont {Maia},
  \citenamefont {Mazalova}, \citenamefont {Meza}, \citenamefont {Michelat},
  \citenamefont {Ourmazd}, \citenamefont {Palmer}, \citenamefont {Ramilli},
  \citenamefont {Schubert}, \citenamefont {Schwander}, \citenamefont {Silenzi},
  \citenamefont {Sztuk-Dambietz}, \citenamefont {Tolstikova}, \citenamefont
  {Chapman}, \citenamefont {Ros}, \citenamefont {Barty}, \citenamefont
  {Fromme}, \citenamefont {Mancuso},\ and\ \citenamefont
  {Schmidt}}]{pandey2019}%
  \BibitemOpen
  \bibfield  {author} {\bibinfo {author} {\bibfnamefont {S.}~\bibnamefont
  {Pandey}}, \bibinfo {author} {\bibfnamefont {R.}~\bibnamefont {Bean}},
  \bibinfo {author} {\bibfnamefont {T.}~\bibnamefont {Sato}}, \bibinfo {author}
  {\bibfnamefont {I.}~\bibnamefont {Poudyal}}, \bibinfo {author} {\bibfnamefont
  {J.}~\bibnamefont {Bielecki}}, \bibinfo {author} {\bibfnamefont
  {J.}~\bibnamefont {Cruz~Villarreal}}, \bibinfo {author} {\bibfnamefont
  {O.}~\bibnamefont {Yefanov}}, \bibinfo {author} {\bibfnamefont
  {V.}~\bibnamefont {Mariani}}, \bibinfo {author} {\bibfnamefont {T.~A.}\
  \bibnamefont {White}}, \bibinfo {author} {\bibfnamefont {C.}~\bibnamefont
  {Kupitz}}, \bibinfo {author} {\bibfnamefont {M.}~\bibnamefont {Hunter}},
  \bibinfo {author} {\bibfnamefont {M.~H.}\ \bibnamefont {Abdellatif}},
  \bibinfo {author} {\bibfnamefont {S.}~\bibnamefont {Bajt}}, \bibinfo {author}
  {\bibfnamefont {V.}~\bibnamefont {Bondar}}, \bibinfo {author} {\bibfnamefont
  {A.}~\bibnamefont {Echelmeier}}, \bibinfo {author} {\bibfnamefont
  {D.}~\bibnamefont {Doppler}}, \bibinfo {author} {\bibfnamefont
  {M.}~\bibnamefont {Emons}}, \bibinfo {author} {\bibfnamefont
  {M.}~\bibnamefont {Frank}}, \bibinfo {author} {\bibfnamefont
  {R.}~\bibnamefont {Fromme}}, \bibinfo {author} {\bibfnamefont
  {Y.}~\bibnamefont {Gevorkov}}, \bibinfo {author} {\bibfnamefont
  {G.}~\bibnamefont {Giovanetti}}, \bibinfo {author} {\bibfnamefont
  {M.}~\bibnamefont {Jiang}}, \bibinfo {author} {\bibfnamefont
  {D.}~\bibnamefont {Kim}}, \bibinfo {author} {\bibfnamefont {Y.}~\bibnamefont
  {Kim}}, \bibinfo {author} {\bibfnamefont {H.}~\bibnamefont {Kirkwood}},
  \bibinfo {author} {\bibfnamefont {A.}~\bibnamefont {Klimovskaia}}, \bibinfo
  {author} {\bibfnamefont {J.}~\bibnamefont {Knoska}}, \bibinfo {author}
  {\bibfnamefont {F.~H.~M.}\ \bibnamefont {Koua}}, \bibinfo {author}
  {\bibfnamefont {R.}~\bibnamefont {Letrun}}, \bibinfo {author} {\bibfnamefont
  {S.}~\bibnamefont {Lisova}}, \bibinfo {author} {\bibfnamefont
  {L.}~\bibnamefont {Maia}}, \bibinfo {author} {\bibfnamefont {V.}~\bibnamefont
  {Mazalova}}, \bibinfo {author} {\bibfnamefont {D.}~\bibnamefont {Meza}},
  \bibinfo {author} {\bibfnamefont {T.}~\bibnamefont {Michelat}}, \bibinfo
  {author} {\bibfnamefont {A.}~\bibnamefont {Ourmazd}}, \bibinfo {author}
  {\bibfnamefont {G.}~\bibnamefont {Palmer}}, \bibinfo {author} {\bibfnamefont
  {M.}~\bibnamefont {Ramilli}}, \bibinfo {author} {\bibfnamefont
  {R.}~\bibnamefont {Schubert}}, \bibinfo {author} {\bibfnamefont
  {P.}~\bibnamefont {Schwander}}, \bibinfo {author} {\bibfnamefont
  {A.}~\bibnamefont {Silenzi}}, \bibinfo {author} {\bibfnamefont
  {J.}~\bibnamefont {Sztuk-Dambietz}}, \bibinfo {author} {\bibfnamefont
  {A.}~\bibnamefont {Tolstikova}}, \bibinfo {author} {\bibfnamefont {H.~N.}\
  \bibnamefont {Chapman}}, \bibinfo {author} {\bibfnamefont {A.}~\bibnamefont
  {Ros}}, \bibinfo {author} {\bibfnamefont {A.}~\bibnamefont {Barty}}, \bibinfo
  {author} {\bibfnamefont {P.}~\bibnamefont {Fromme}}, \bibinfo {author}
  {\bibfnamefont {A.~P.}\ \bibnamefont {Mancuso}}, \ and\ \bibinfo {author}
  {\bibfnamefont {M.}~\bibnamefont {Schmidt}},\ }\bibfield  {title} {\enquote
  {\bibinfo {title} {Time-resolved serial femtosecond crystallography at the
  {European XFEL}},}\ }\href {https://doi.org/10.1038/s41592-019-0628-z}
  {\bibfield  {journal} {\bibinfo  {journal} {Nature Methods}\ } (\bibinfo
  {year} {2019})}\BibitemShut {NoStop}%
\bibitem [{\citenamefont {Behrens}\ \emph {et~al.}(2012)\citenamefont
  {Behrens}, \citenamefont {Gerasimova}, \citenamefont {Gerth}, \citenamefont
  {Schmidt}, \citenamefont {Schneidmiller}, \citenamefont {Serkez},
  \citenamefont {Wesch},\ and\ \citenamefont {Yurkov}}]{behrens12}%
  \BibitemOpen
  \bibfield  {author} {\bibinfo {author} {\bibfnamefont {C.}~\bibnamefont
  {Behrens}}, \bibinfo {author} {\bibfnamefont {N.}~\bibnamefont {Gerasimova}},
  \bibinfo {author} {\bibfnamefont {C.}~\bibnamefont {Gerth}}, \bibinfo
  {author} {\bibfnamefont {B.}~\bibnamefont {Schmidt}}, \bibinfo {author}
  {\bibfnamefont {E.~A.}\ \bibnamefont {Schneidmiller}}, \bibinfo {author}
  {\bibfnamefont {S.}~\bibnamefont {Serkez}}, \bibinfo {author} {\bibfnamefont
  {S.}~\bibnamefont {Wesch}}, \ and\ \bibinfo {author} {\bibfnamefont {M.~V.}\
  \bibnamefont {Yurkov}},\ }\bibfield  {title} {\enquote {\bibinfo {title}
  {Constraints on photon pulse duration from longitudinal electron beam
  diagnostics at a soft x-ray free-electron laser},}\ }\href {\doibase
  10.1103/PhysRevSTAB.15.030707} {\bibfield  {journal} {\bibinfo  {journal}
  {Phys. Rev. ST Accel. Beams}\ }\textbf {\bibinfo {volume} {15}},\ \bibinfo
  {pages} {030707} (\bibinfo {year} {2012})}\BibitemShut {NoStop}%
\bibitem [{\citenamefont {Behrens}\ \emph {et~al.}(2014)\citenamefont
  {Behrens}, \citenamefont {Decker}, \citenamefont {Ding}, \citenamefont
  {Dolgashev}, \citenamefont {Frisch}, \citenamefont {Huang}, \citenamefont
  {Krejcik}, \citenamefont {Loos}, \citenamefont {Lutman}, \citenamefont
  {Maxwell}, \citenamefont {Turner}, \citenamefont {Wang}, \citenamefont
  {Wang}, \citenamefont {Welch},\ and\ \citenamefont {Wu}}]{behrens14}%
  \BibitemOpen
  \bibfield  {author} {\bibinfo {author} {\bibfnamefont {C.}~\bibnamefont
  {Behrens}}, \bibinfo {author} {\bibfnamefont {F.-J.}\ \bibnamefont {Decker}},
  \bibinfo {author} {\bibfnamefont {Y.}~\bibnamefont {Ding}}, \bibinfo {author}
  {\bibfnamefont {V.~A.}\ \bibnamefont {Dolgashev}}, \bibinfo {author}
  {\bibfnamefont {J.}~\bibnamefont {Frisch}}, \bibinfo {author} {\bibfnamefont
  {Z.}~\bibnamefont {Huang}}, \bibinfo {author} {\bibfnamefont
  {P.}~\bibnamefont {Krejcik}}, \bibinfo {author} {\bibfnamefont
  {H.}~\bibnamefont {Loos}}, \bibinfo {author} {\bibfnamefont {A.}~\bibnamefont
  {Lutman}}, \bibinfo {author} {\bibfnamefont {T.~J.}\ \bibnamefont {Maxwell}},
  \bibinfo {author} {\bibfnamefont {J.}~\bibnamefont {Turner}}, \bibinfo
  {author} {\bibfnamefont {J.}~\bibnamefont {Wang}}, \bibinfo {author}
  {\bibfnamefont {M.-H.}\ \bibnamefont {Wang}}, \bibinfo {author}
  {\bibfnamefont {J.}~\bibnamefont {Welch}}, \ and\ \bibinfo {author}
  {\bibfnamefont {J.}~\bibnamefont {Wu}},\ }\bibfield  {title} {\enquote
  {\bibinfo {title} {Few-femtosecond time-resolved measurements of x-ray
  free-electron lasers},}\ }\href {https://doi.org/10.1038/ncomms4762}
  {\bibfield  {journal} {\bibinfo  {journal} {Nature Communications}\ }\textbf
  {\bibinfo {volume} {5}},\ \bibinfo {pages} {3762} (\bibinfo {year}
  {2014})}\BibitemShut {NoStop}%
\bibitem [{\citenamefont {Zagorodnov}, \citenamefont {Dohlus},\ and\
  \citenamefont {Tomin}(2019)}]{Zagorodnov2019}%
  \BibitemOpen
  \bibfield  {author} {\bibinfo {author} {\bibfnamefont {I.}~\bibnamefont
  {Zagorodnov}}, \bibinfo {author} {\bibfnamefont {M.}~\bibnamefont {Dohlus}},
  \ and\ \bibinfo {author} {\bibfnamefont {S.}~\bibnamefont {Tomin}},\
  }\bibfield  {title} {\enquote {\bibinfo {title} {Accelerator beam dynamics at
  the european x-ray free electron laser},}\ }\href {\doibase
  10.1103/PhysRevAccelBeams.22.024401} {\bibfield  {journal} {\bibinfo
  {journal} {Phys. Rev. Accel. Beams}\ }\textbf {\bibinfo {volume} {22}},\
  \bibinfo {pages} {024401} (\bibinfo {year} {2019})}\BibitemShut {NoStop}%
\bibitem [{\citenamefont {Rota}\ \emph {et~al.}(2019)\citenamefont {Rota},
  \citenamefont {Caselle}, \citenamefont {Bründermann}, \citenamefont
  {Funkner}, \citenamefont {Gerth}, \citenamefont {Kehrer}, \citenamefont
  {Mielczarek}, \citenamefont {Makowski}, \citenamefont {Mozzanica},
  \citenamefont {Müller}, \citenamefont {Nasse}, \citenamefont {Niehues},
  \citenamefont {Patil}, \citenamefont {Schmitt}, \citenamefont {Schönfeldt},
  \citenamefont {Steffen},\ and\ \citenamefont {Weber}}]{rota2019}%
  \BibitemOpen
  \bibfield  {author} {\bibinfo {author} {\bibfnamefont {L.}~\bibnamefont
  {Rota}}, \bibinfo {author} {\bibfnamefont {M.}~\bibnamefont {Caselle}},
  \bibinfo {author} {\bibfnamefont {E.}~\bibnamefont {Bründermann}}, \bibinfo
  {author} {\bibfnamefont {S.}~\bibnamefont {Funkner}}, \bibinfo {author}
  {\bibfnamefont {C.}~\bibnamefont {Gerth}}, \bibinfo {author} {\bibfnamefont
  {B.}~\bibnamefont {Kehrer}}, \bibinfo {author} {\bibfnamefont
  {A.}~\bibnamefont {Mielczarek}}, \bibinfo {author} {\bibfnamefont
  {D.}~\bibnamefont {Makowski}}, \bibinfo {author} {\bibfnamefont
  {A.}~\bibnamefont {Mozzanica}}, \bibinfo {author} {\bibfnamefont {A.-S.}\
  \bibnamefont {Müller}}, \bibinfo {author} {\bibfnamefont {M.~J.}\
  \bibnamefont {Nasse}}, \bibinfo {author} {\bibfnamefont {G.}~\bibnamefont
  {Niehues}}, \bibinfo {author} {\bibfnamefont {M.}~\bibnamefont {Patil}},
  \bibinfo {author} {\bibfnamefont {B.}~\bibnamefont {Schmitt}}, \bibinfo
  {author} {\bibfnamefont {P.}~\bibnamefont {Schönfeldt}}, \bibinfo {author}
  {\bibfnamefont {B.}~\bibnamefont {Steffen}}, \ and\ \bibinfo {author}
  {\bibfnamefont {M.}~\bibnamefont {Weber}},\ }\bibfield  {title} {\enquote
  {\bibinfo {title} {{KALYPSO:} linear array detector for high-repetition rate
  and real-time beam diagnostics},}\ }\href {\doibase
  https://doi.org/10.1016/j.nima.2018.10.093} {\bibfield  {journal} {\bibinfo
  {journal} {Nuclear Instruments and Methods in Physics Research Section A:
  Accelerators, Spectrometers, Detectors and Associated Equipment}\ }\textbf
  {\bibinfo {volume} {936}},\ \bibinfo {pages} {10 -- 13} (\bibinfo {year}
  {2019})}\BibitemShut {NoStop}%
\bibitem [{\citenamefont {Steffen}(2007)}]{berndthesis}%
  \BibitemOpen
  \bibfield  {author} {\bibinfo {author} {\bibfnamefont {B.}~\bibnamefont
  {Steffen}},\ }\emph {\bibinfo {title} {Electro-optic methods for bunch length
  diagnostics at {FLASH}}},\ \href@noop {} {Ph.D. thesis},\ \bibinfo  {school}
  {University of Hamburg} (\bibinfo {year} {2007})\BibitemShut {NoStop}%
\bibitem [{\citenamefont {Wu}\ and\ \citenamefont {Zhang}(1997)}]{wu97}%
  \BibitemOpen
  \bibfield  {author} {\bibinfo {author} {\bibfnamefont {Q.}~\bibnamefont
  {Wu}}\ and\ \bibinfo {author} {\bibfnamefont {X.-C.}\ \bibnamefont {Zhang}},\
  }\bibfield  {title} {\enquote {\bibinfo {title} {7 terahertz broadband {GaP}
  electro-optic sensor},}\ }\href@noop {} {\bibfield  {journal} {\bibinfo
  {journal} {Appl. Phys. Lett.}\ }\textbf {\bibinfo {volume} {70}},\ \bibinfo
  {pages} {14} (\bibinfo {year} {1997})}\BibitemShut {NoStop}%
\bibitem [{\citenamefont {Casalbuoni}\ \emph {et~al.}(2008)\citenamefont
  {Casalbuoni}, \citenamefont {Schlarb}, \citenamefont {Schmidt}, \citenamefont
  {Schm\"user}, \citenamefont {Steffen},\ and\ \citenamefont
  {Winter}}]{casalbuoni08}%
  \BibitemOpen
  \bibfield  {author} {\bibinfo {author} {\bibfnamefont {S.}~\bibnamefont
  {Casalbuoni}}, \bibinfo {author} {\bibfnamefont {H.}~\bibnamefont {Schlarb}},
  \bibinfo {author} {\bibfnamefont {B.}~\bibnamefont {Schmidt}}, \bibinfo
  {author} {\bibfnamefont {P.}~\bibnamefont {Schm\"user}}, \bibinfo {author}
  {\bibfnamefont {B.}~\bibnamefont {Steffen}}, \ and\ \bibinfo {author}
  {\bibfnamefont {A.}~\bibnamefont {Winter}},\ }\bibfield  {title} {\enquote
  {\bibinfo {title} {Numerical studies on the electro-optic detection of
  femtosecond electron bunches},}\ }\href {\doibase
  10.1103/PhysRevSTAB.11.072802} {\bibfield  {journal} {\bibinfo  {journal}
  {Phys. Rev. ST Accel. Beams}\ }\textbf {\bibinfo {volume} {11}},\ \bibinfo
  {pages} {072802} (\bibinfo {year} {2008})}\BibitemShut {NoStop}%
\bibitem [{\citenamefont {Paradis}\ \emph {et~al.}(2018)\citenamefont
  {Paradis}, \citenamefont {Drs}, \citenamefont {Modsching}, \citenamefont
  {Razskazovskaya}, \citenamefont {Meyer}, \citenamefont {Kr\"{a}nkel},
  \citenamefont {Saraceno}, \citenamefont {Wittwer},\ and\ \citenamefont
  {S\"{u}dmeyer}}]{paradis2018}%
  \BibitemOpen
  \bibfield  {author} {\bibinfo {author} {\bibfnamefont {C.}~\bibnamefont
  {Paradis}}, \bibinfo {author} {\bibfnamefont {J.}~\bibnamefont {Drs}},
  \bibinfo {author} {\bibfnamefont {N.}~\bibnamefont {Modsching}}, \bibinfo
  {author} {\bibfnamefont {O.}~\bibnamefont {Razskazovskaya}}, \bibinfo
  {author} {\bibfnamefont {F.}~\bibnamefont {Meyer}}, \bibinfo {author}
  {\bibfnamefont {C.}~\bibnamefont {Kr\"{a}nkel}}, \bibinfo {author}
  {\bibfnamefont {C.~J.}\ \bibnamefont {Saraceno}}, \bibinfo {author}
  {\bibfnamefont {V.~J.}\ \bibnamefont {Wittwer}}, \ and\ \bibinfo {author}
  {\bibfnamefont {T.}~\bibnamefont {S\"{u}dmeyer}},\ }\bibfield  {title}
  {\enquote {\bibinfo {title} {Broadband terahertz pulse generation driven by
  an ultrafast thin-disk laser oscillator},}\ }\href {\doibase
  10.1364/OE.26.026377} {\bibfield  {journal} {\bibinfo  {journal} {Opt.
  Express}\ }\textbf {\bibinfo {volume} {26}},\ \bibinfo {pages} {26377--26384}
  (\bibinfo {year} {2018})}\BibitemShut {NoStop}%
\bibitem [{\citenamefont {Leitenstorfer}\ \emph {et~al.}(1999)\citenamefont
  {Leitenstorfer}, \citenamefont {Hunsche}, \citenamefont {Shah}, \citenamefont
  {Nuss},\ and\ \citenamefont {Knox}}]{leitenstorfer99}%
  \BibitemOpen
  \bibfield  {author} {\bibinfo {author} {\bibfnamefont {A.}~\bibnamefont
  {Leitenstorfer}}, \bibinfo {author} {\bibfnamefont {S.}~\bibnamefont
  {Hunsche}}, \bibinfo {author} {\bibfnamefont {J.}~\bibnamefont {Shah}},
  \bibinfo {author} {\bibfnamefont {M.~C.}\ \bibnamefont {Nuss}}, \ and\
  \bibinfo {author} {\bibfnamefont {W.~H.}\ \bibnamefont {Knox}},\ }\bibfield
  {title} {\enquote {\bibinfo {title} {Detectors and sources for ultrabroadband
  electro-optic sampling: Experiment and theory},}\ }\href@noop {} {\bibfield
  {journal} {\bibinfo  {journal} {Appl. Phys. Lett.}\ }\textbf {\bibinfo
  {volume} {74}},\ \bibinfo {pages} {1516} (\bibinfo {year}
  {1999})}\BibitemShut {NoStop}%
\bibitem [{\citenamefont {{PCI Industrial Computer Manufacturers Group
  (PICMIC)}}(2019)}]{mtca_picmic}%
  \BibitemOpen
  \bibfield  {author} {\bibinfo {author} {\bibnamefont {{PCI Industrial
  Computer Manufacturers Group (PICMIC)}}},\ }\href
  {https://www.picmg.org/openstandards/microtca/} {} (\bibinfo {year}
  {2019})\BibitemShut {NoStop}%
\bibitem [{\citenamefont {M\"uller}\ \emph {et~al.}(2009)\citenamefont
  {M\"uller}, \citenamefont {Hunziker}, \citenamefont {Schlott}, \citenamefont
  {Steffen}, \citenamefont {Treyer},\ and\ \citenamefont {Feurer}}]{mueller09}%
  \BibitemOpen
  \bibfield  {author} {\bibinfo {author} {\bibfnamefont {F.}~\bibnamefont
  {M\"uller}}, \bibinfo {author} {\bibfnamefont {S.}~\bibnamefont {Hunziker}},
  \bibinfo {author} {\bibfnamefont {V.}~\bibnamefont {Schlott}}, \bibinfo
  {author} {\bibfnamefont {B.}~\bibnamefont {Steffen}}, \bibinfo {author}
  {\bibfnamefont {D.}~\bibnamefont {Treyer}}, \ and\ \bibinfo {author}
  {\bibfnamefont {T.}~\bibnamefont {Feurer}},\ }\bibfield  {title} {\enquote
  {\bibinfo {title} {Ytterbium fiber laser for electro-optical pulse length
  measurements at the {SwissFEL}},}\ }in\ \href@noop {} {\emph {\bibinfo
  {booktitle} {Proceedings of DIPAC2009}}}\ (\bibinfo {address} {Basel,
  Switzerland},\ \bibinfo {year} {2009})\BibitemShut {NoStop}%
\bibitem [{\citenamefont {M\"uller}(2011)}]{felixthesis}%
  \BibitemOpen
  \bibfield  {author} {\bibinfo {author} {\bibfnamefont {F.}~\bibnamefont
  {M\"uller}},\ }\emph {\bibinfo {title} {Electro-Optical Bunch Length
  Measurements at the {Swiss Light Source}}},\ \href@noop {} {Ph.D. thesis},\
  \bibinfo  {school} {University of Bern}, \bibinfo {address} {Bern,
  Switzerland} (\bibinfo {year} {2011})\BibitemShut {NoStop}%
\bibitem [{\citenamefont {Hofer}\ \emph {et~al.}(1991)\citenamefont {Hofer},
  \citenamefont {Fermann}, \citenamefont {Haberl}, \citenamefont {Ober},\ and\
  \citenamefont {Schmidt}}]{Hofer91}%
  \BibitemOpen
  \bibfield  {author} {\bibinfo {author} {\bibfnamefont {M.}~\bibnamefont
  {Hofer}}, \bibinfo {author} {\bibfnamefont {M.~E.}\ \bibnamefont {Fermann}},
  \bibinfo {author} {\bibfnamefont {F.}~\bibnamefont {Haberl}}, \bibinfo
  {author} {\bibfnamefont {M.~H.}\ \bibnamefont {Ober}}, \ and\ \bibinfo
  {author} {\bibfnamefont {A.~J.}\ \bibnamefont {Schmidt}},\ }\bibfield
  {title} {\enquote {\bibinfo {title} {Mode locking with cross-phase and
  self-phase modulation},}\ }\href {\doibase 10.1364/OL.16.000502} {\bibfield
  {journal} {\bibinfo  {journal} {Opt. Lett.}\ }\textbf {\bibinfo {volume}
  {16}},\ \bibinfo {pages} {502--504} (\bibinfo {year} {1991})}\BibitemShut
  {NoStop}%
\bibitem [{\citenamefont {Hidv{\'e}gi}\ \emph {et~al.}(2014)\citenamefont
  {Hidv{\'e}gi}, \citenamefont {Ge{\ss}ler}, \citenamefont {Kay}, \citenamefont
  {Petrosyan}, \citenamefont {Petrosyan}, \citenamefont {Petrosyan},
  \citenamefont {Aghababyan}, \citenamefont {Stechmann}, \citenamefont
  {Rehlich},\ and\ \citenamefont {Bohm}}]{x2Timer}%
  \BibitemOpen
  \bibfield  {author} {\bibinfo {author} {\bibfnamefont {A.}~\bibnamefont
  {Hidv{\'e}gi}}, \bibinfo {author} {\bibfnamefont {P.}~\bibnamefont
  {Ge{\ss}ler}}, \bibinfo {author} {\bibfnamefont {H.}~\bibnamefont {Kay}},
  \bibinfo {author} {\bibfnamefont {V.}~\bibnamefont {Petrosyan}}, \bibinfo
  {author} {\bibfnamefont {G.}~\bibnamefont {Petrosyan}}, \bibinfo {author}
  {\bibfnamefont {L.}~\bibnamefont {Petrosyan}}, \bibinfo {author}
  {\bibfnamefont {A.}~\bibnamefont {Aghababyan}}, \bibinfo {author}
  {\bibfnamefont {C.}~\bibnamefont {Stechmann}}, \bibinfo {author}
  {\bibfnamefont {K.}~\bibnamefont {Rehlich}}, \ and\ \bibinfo {author}
  {\bibfnamefont {C.}~\bibnamefont {Bohm}},\ }\bibfield  {title} {\enquote
  {\bibinfo {title} {A trigger fanout {Rear-Transition Module} for the
  {European XFEL} timing system},}\ }in\ \href {\doibase
  10.1109/RTC.2014.7097423} {\emph {\bibinfo {booktitle} {Proceedings of the
  19th IEEE-NPSS Real Time Conference}}}\ (\bibinfo {address} {Nara, Japan},\
  \bibinfo {year} {2014})\BibitemShut {NoStop}%
\bibitem [{\citenamefont {Azima}\ \emph {et~al.}(2006)\citenamefont {Azima},
  \citenamefont {D\"usterer}, \citenamefont {Schlarb}, \citenamefont
  {Feldhaus}, \citenamefont {Cavalieri}, \citenamefont {Fritz},\ and\
  \citenamefont {Sengstock}}]{azima06}%
  \BibitemOpen
  \bibfield  {author} {\bibinfo {author} {\bibfnamefont {A.}~\bibnamefont
  {Azima}}, \bibinfo {author} {\bibfnamefont {S.}~\bibnamefont {D\"usterer}},
  \bibinfo {author} {\bibfnamefont {H.}~\bibnamefont {Schlarb}}, \bibinfo
  {author} {\bibfnamefont {J.}~\bibnamefont {Feldhaus}}, \bibinfo {author}
  {\bibfnamefont {A.}~\bibnamefont {Cavalieri}}, \bibinfo {author}
  {\bibfnamefont {D.}~\bibnamefont {Fritz}}, \ and\ \bibinfo {author}
  {\bibfnamefont {K.}~\bibnamefont {Sengstock}},\ }\bibfield  {title} {\enquote
  {\bibinfo {title} {Jitter measurements by spatial electro-optical sampling at
  the {FLASH} free electron laser},}\ }in\ \href@noop {} {\emph {\bibinfo
  {booktitle} {Proceedings of the EPAC 2006}}}\ (\bibinfo {address} {Edinburgh,
  Scotland},\ \bibinfo {year} {2006})\BibitemShut {NoStop}%
\bibitem [{\citenamefont {Doolittle}, \citenamefont {Champion},\ and\
  \citenamefont {Ma.}(2006)}]{doolittle2006}%
  \BibitemOpen
  \bibfield  {author} {\bibinfo {author} {\bibfnamefont {L.~R.}\ \bibnamefont
  {Doolittle}}, \bibinfo {author} {\bibfnamefont {M.~S.}\ \bibnamefont
  {Champion}}, \ and\ \bibinfo {author} {\bibfnamefont {H.}~\bibnamefont
  {Ma.}},\ }\bibfield  {title} {{\selectlanguage {english}\enquote {\bibinfo
  {title} {Digital low-level rf control using {Non-IQ} sampling},}\ }}in\ \href
  {\doibase 10.1.1.585.1925} {{\selectlanguage {english}\emph {\bibinfo
  {booktitle} {Proc. of Linear Accelerator Conference (LINAC'06)}}}},\ \bibinfo
  {series and number} {\bibinfo {series} {Linear Accelerator Conference}\
  No.~\bibinfo {number} {23}}\ (\bibinfo  {publisher} {JACoW},\ \bibinfo
  {address} {Geneva, Switzerland},\ \bibinfo {year} {2006})\ pp.\ \bibinfo
  {pages} {568--570}\BibitemShut {NoStop}%
\bibitem [{\citenamefont {{Przygoda}}\ \emph {et~al.}(2017)\citenamefont
  {{Przygoda}}, \citenamefont {{Rybaniec}}, \citenamefont {{Butkowski}},
  \citenamefont {{Gerth}}, \citenamefont {{Peier}}, \citenamefont {{Schmidt}},
  \citenamefont {{Steffen}},\ and\ \citenamefont {{Schlarb}}}]{przygoda2017}%
  \BibitemOpen
  \bibfield  {author} {\bibinfo {author} {\bibfnamefont {K.}~\bibnamefont
  {{Przygoda}}}, \bibinfo {author} {\bibfnamefont {R.}~\bibnamefont
  {{Rybaniec}}}, \bibinfo {author} {\bibfnamefont {.}~\bibnamefont
  {{Butkowski}}}, \bibinfo {author} {\bibfnamefont {C.}~\bibnamefont
  {{Gerth}}}, \bibinfo {author} {\bibfnamefont {P.}~\bibnamefont {{Peier}}},
  \bibinfo {author} {\bibfnamefont {C.}~\bibnamefont {{Schmidt}}}, \bibinfo
  {author} {\bibfnamefont {B.}~\bibnamefont {{Steffen}}}, \ and\ \bibinfo
  {author} {\bibfnamefont {H.}~\bibnamefont {{Schlarb}}},\ }\bibfield  {title}
  {\enquote {\bibinfo {title} {{MicroTCA.4}-based {RF} and laser cavities
  regulation including piezocontrols},}\ }\href {\doibase
  10.1109/TNS.2017.2710286} {\bibfield  {journal} {\bibinfo  {journal} {IEEE
  Transactions on Nuclear Science}\ }\textbf {\bibinfo {volume} {64}},\
  \bibinfo {pages} {1389--1394} (\bibinfo {year} {2017})}\BibitemShut {NoStop}%
\bibitem [{\citenamefont {Gerth}\ \emph {et~al.}(2019)\citenamefont {Gerth},
  \citenamefont {Brenner}, \citenamefont {Caselle}, \citenamefont
  {D{\"{u}}sterer}, \citenamefont {Haack}, \citenamefont {Makowski},
  \citenamefont {Mielczarek}, \citenamefont {Palutke}, \citenamefont {Rota},
  \citenamefont {Rybnikov}, \citenamefont {Schmidt}, \citenamefont {Steffen},\
  and\ \citenamefont {Tiedtke}}]{gerth19}%
  \BibitemOpen
  \bibfield  {author} {\bibinfo {author} {\bibfnamefont {C.}~\bibnamefont
  {Gerth}}, \bibinfo {author} {\bibfnamefont {G.}~\bibnamefont {Brenner}},
  \bibinfo {author} {\bibfnamefont {M.}~\bibnamefont {Caselle}}, \bibinfo
  {author} {\bibfnamefont {S.}~\bibnamefont {D{\"{u}}sterer}}, \bibinfo
  {author} {\bibfnamefont {D.}~\bibnamefont {Haack}}, \bibinfo {author}
  {\bibfnamefont {D.}~\bibnamefont {Makowski}}, \bibinfo {author}
  {\bibfnamefont {A.}~\bibnamefont {Mielczarek}}, \bibinfo {author}
  {\bibfnamefont {S.}~\bibnamefont {Palutke}}, \bibinfo {author} {\bibfnamefont
  {L.}~\bibnamefont {Rota}}, \bibinfo {author} {\bibfnamefont {V.}~\bibnamefont
  {Rybnikov}}, \bibinfo {author} {\bibfnamefont {C.}~\bibnamefont {Schmidt}},
  \bibinfo {author} {\bibfnamefont {B.}~\bibnamefont {Steffen}}, \ and\
  \bibinfo {author} {\bibfnamefont {K.}~\bibnamefont {Tiedtke}},\ }\bibfield
  {title} {\enquote {\bibinfo {title} {{Linear array detector for online
  diagnostics of spectral distributions at MHz repetition rates}},}\ }\href
  {\doibase 10.1107/S1600577519007835} {\bibfield  {journal} {\bibinfo
  {journal} {Journal of Synchrotron Radiation}\ }\textbf {\bibinfo {volume}
  {26}},\ \bibinfo {pages} {1514--1522} (\bibinfo {year} {2019})}\BibitemShut
  {NoStop}%
\bibitem [{\citenamefont {Mozzanica}\ \emph {et~al.}(2012)\citenamefont
  {Mozzanica}, \citenamefont {Bergamaschi}, \citenamefont {Dinapoli},
  \citenamefont {Graafsma}, \citenamefont {Greiffenberg}, \citenamefont
  {Henrich}, \citenamefont {Johnson}, \citenamefont {Lohmann}, \citenamefont
  {Valeria}, \citenamefont {Schmitt},\ and\ \citenamefont
  {Xintian}}]{mozzanica12}%
  \BibitemOpen
  \bibfield  {author} {\bibinfo {author} {\bibfnamefont {A.}~\bibnamefont
  {Mozzanica}}, \bibinfo {author} {\bibfnamefont {A.}~\bibnamefont
  {Bergamaschi}}, \bibinfo {author} {\bibfnamefont {R.}~\bibnamefont
  {Dinapoli}}, \bibinfo {author} {\bibfnamefont {H.}~\bibnamefont {Graafsma}},
  \bibinfo {author} {\bibfnamefont {D.}~\bibnamefont {Greiffenberg}}, \bibinfo
  {author} {\bibfnamefont {B.}~\bibnamefont {Henrich}}, \bibinfo {author}
  {\bibfnamefont {I.}~\bibnamefont {Johnson}}, \bibinfo {author} {\bibfnamefont
  {M.}~\bibnamefont {Lohmann}}, \bibinfo {author} {\bibfnamefont
  {R.}~\bibnamefont {Valeria}}, \bibinfo {author} {\bibfnamefont
  {B.}~\bibnamefont {Schmitt}}, \ and\ \bibinfo {author} {\bibfnamefont
  {S.}~\bibnamefont {Xintian}},\ }\bibfield  {title} {\enquote {\bibinfo
  {title} {The {GOTTHARD} charge integrating readout detector: design and
  characterization},}\ }\href {http://stacks.iop.org/1748-0221/7/i=01/a=C01019}
  {\bibfield  {journal} {\bibinfo  {journal} {Journal of Instrumentation}\
  }\textbf {\bibinfo {volume} {7}},\ \bibinfo {pages} {C01019} (\bibinfo {year}
  {2012})}\BibitemShut {NoStop}%
\bibitem [{\citenamefont {Viti}\ \emph {et~al.}(2018)\citenamefont {Viti},
  \citenamefont {Czwalinna}, \citenamefont {Dinter}, \citenamefont {Gerth},
  \citenamefont {Przygoda}, \citenamefont {Rybaniec},\ and\ \citenamefont
  {Schlarb}}]{viti2017}%
  \BibitemOpen
  \bibfield  {author} {\bibinfo {author} {\bibfnamefont {M.}~\bibnamefont
  {Viti}}, \bibinfo {author} {\bibfnamefont {M.~K.}\ \bibnamefont {Czwalinna}},
  \bibinfo {author} {\bibfnamefont {H.}~\bibnamefont {Dinter}}, \bibinfo
  {author} {\bibfnamefont {C.}~\bibnamefont {Gerth}}, \bibinfo {author}
  {\bibfnamefont {K.}~\bibnamefont {Przygoda}}, \bibinfo {author}
  {\bibfnamefont {R.}~\bibnamefont {Rybaniec}}, \ and\ \bibinfo {author}
  {\bibfnamefont {H.}~\bibnamefont {Schlarb}},\ }\bibfield  {title}
  {{\selectlanguage {english}\enquote {\bibinfo {title} {The bunch arrival time
  monitor at {FLASH} and {E}uropean {XFEL}},}\ }}in\ \href {\doibase
  https://doi.org/10.18429/JACoW-ICALEPCS2017-TUPHA125} {{\selectlanguage
  {english}\emph {\bibinfo {booktitle} {Proc. of ICALEPCS 2017}}}},\ \bibinfo
  {series and number} {\bibinfo {series} {International Conference on
  Accelerator and Large Experimental Control Systems}\ No.~\bibinfo {number}
  {16}}\ (\bibinfo  {publisher} {JACoW},\ \bibinfo {address} {Geneva,
  Switzerland},\ \bibinfo {year} {2018})\ pp.\ \bibinfo {pages}
  {701--705}\BibitemShut {NoStop}%
\bibitem [{\citenamefont {Kirkwood}\ \emph {et~al.}(2019)\citenamefont
  {Kirkwood}, \citenamefont {Letrun}, \citenamefont {Tanikawa}, \citenamefont
  {Liu}, \citenamefont {Nakatsutsumi}, \citenamefont {Emons}, \citenamefont
  {Jezynski}, \citenamefont {Palmer}, \citenamefont {Lederer}, \citenamefont
  {Bean}, \citenamefont {Buck}, \citenamefont {Cafisio}, \citenamefont
  {Graceffa}, \citenamefont {Gr\"{u}nert}, \citenamefont {G\"{o}de},
  \citenamefont {H\"{o}ppner}, \citenamefont {Kim}, \citenamefont {Konopkova},
  \citenamefont {Mills}, \citenamefont {Makita}, \citenamefont {Pelka},
  \citenamefont {Preston}, \citenamefont {Sikorski}, \citenamefont {Takem},
  \citenamefont {Giewekemeyer}, \citenamefont {Chollet}, \citenamefont
  {Vagovic}, \citenamefont {Chapman}, \citenamefont {Mancuso},\ and\
  \citenamefont {Sato}}]{kirkwood2019}%
  \BibitemOpen
  \bibfield  {author} {\bibinfo {author} {\bibfnamefont {H.~J.}\ \bibnamefont
  {Kirkwood}}, \bibinfo {author} {\bibfnamefont {R.}~\bibnamefont {Letrun}},
  \bibinfo {author} {\bibfnamefont {T.}~\bibnamefont {Tanikawa}}, \bibinfo
  {author} {\bibfnamefont {J.}~\bibnamefont {Liu}}, \bibinfo {author}
  {\bibfnamefont {M.}~\bibnamefont {Nakatsutsumi}}, \bibinfo {author}
  {\bibfnamefont {M.}~\bibnamefont {Emons}}, \bibinfo {author} {\bibfnamefont
  {T.}~\bibnamefont {Jezynski}}, \bibinfo {author} {\bibfnamefont
  {G.}~\bibnamefont {Palmer}}, \bibinfo {author} {\bibfnamefont
  {M.}~\bibnamefont {Lederer}}, \bibinfo {author} {\bibfnamefont
  {R.}~\bibnamefont {Bean}}, \bibinfo {author} {\bibfnamefont {J.}~\bibnamefont
  {Buck}}, \bibinfo {author} {\bibfnamefont {S.~D.~D.}\ \bibnamefont
  {Cafisio}}, \bibinfo {author} {\bibfnamefont {R.}~\bibnamefont {Graceffa}},
  \bibinfo {author} {\bibfnamefont {J.}~\bibnamefont {Gr\"{u}nert}}, \bibinfo
  {author} {\bibfnamefont {S.}~\bibnamefont {G\"{o}de}}, \bibinfo {author}
  {\bibfnamefont {H.}~\bibnamefont {H\"{o}ppner}}, \bibinfo {author}
  {\bibfnamefont {Y.}~\bibnamefont {Kim}}, \bibinfo {author} {\bibfnamefont
  {Z.}~\bibnamefont {Konopkova}}, \bibinfo {author} {\bibfnamefont
  {G.}~\bibnamefont {Mills}}, \bibinfo {author} {\bibfnamefont
  {M.}~\bibnamefont {Makita}}, \bibinfo {author} {\bibfnamefont
  {A.}~\bibnamefont {Pelka}}, \bibinfo {author} {\bibfnamefont {T.~R.}\
  \bibnamefont {Preston}}, \bibinfo {author} {\bibfnamefont {M.}~\bibnamefont
  {Sikorski}}, \bibinfo {author} {\bibfnamefont {C.~M.~S.}\ \bibnamefont
  {Takem}}, \bibinfo {author} {\bibfnamefont {K.}~\bibnamefont {Giewekemeyer}},
  \bibinfo {author} {\bibfnamefont {M.}~\bibnamefont {Chollet}}, \bibinfo
  {author} {\bibfnamefont {P.}~\bibnamefont {Vagovic}}, \bibinfo {author}
  {\bibfnamefont {H.~N.}\ \bibnamefont {Chapman}}, \bibinfo {author}
  {\bibfnamefont {A.~P.}\ \bibnamefont {Mancuso}}, \ and\ \bibinfo {author}
  {\bibfnamefont {T.}~\bibnamefont {Sato}},\ }\bibfield  {title} {\enquote
  {\bibinfo {title} {Initial observations of the femtosecond timing jitter at
  the {European XFEL}},}\ }\href {\doibase 10.1364/OL.44.001650} {\bibfield
  {journal} {\bibinfo  {journal} {Optics Letters}\ }\textbf {\bibinfo {volume}
  {44}},\ \bibinfo {pages} {1650--1653} (\bibinfo {year} {2019})}\BibitemShut
  {NoStop}%
\bibitem [{\citenamefont {Schulz}\ \emph {et~al.}(2019)\citenamefont {Schulz},
  \citenamefont {Czwalinna}, \citenamefont {Felber}, \citenamefont {Fenner},
  \citenamefont {Gerth}, \citenamefont {Kozak}, \citenamefont {Lamb},
  \citenamefont {Lautenschlager}, \citenamefont {Ludwig}, \citenamefont
  {Mavric}, \citenamefont {Müller}, \citenamefont {Pfeiffer}, \citenamefont
  {Schlarb}, \citenamefont {M.~Titberidze}, \citenamefont {Zummack},\ and\
  \citenamefont {Sydlo}}]{schulz2019}%
  \BibitemOpen
  \bibfield  {author} {\bibinfo {author} {\bibfnamefont {S.}~\bibnamefont
  {Schulz}}, \bibinfo {author} {\bibfnamefont {M.~K.}\ \bibnamefont
  {Czwalinna}}, \bibinfo {author} {\bibfnamefont {M.}~\bibnamefont {Felber}},
  \bibinfo {author} {\bibfnamefont {M.}~\bibnamefont {Fenner}}, \bibinfo
  {author} {\bibfnamefont {C.}~\bibnamefont {Gerth}}, \bibinfo {author}
  {\bibfnamefont {T.}~\bibnamefont {Kozak}}, \bibinfo {author} {\bibfnamefont
  {T.}~\bibnamefont {Lamb}}, \bibinfo {author} {\bibfnamefont {B.}~\bibnamefont
  {Lautenschlager}}, \bibinfo {author} {\bibfnamefont {F.}~\bibnamefont
  {Ludwig}}, \bibinfo {author} {\bibfnamefont {U.}~\bibnamefont {Mavric}},
  \bibinfo {author} {\bibfnamefont {J.}~\bibnamefont {Müller}}, \bibinfo
  {author} {\bibfnamefont {S.}~\bibnamefont {Pfeiffer}}, \bibinfo {author}
  {\bibfnamefont {H.}~\bibnamefont {Schlarb}}, \bibinfo {author} {\bibfnamefont
  {C.~S.}\ \bibnamefont {M.~Titberidze}}, \bibinfo {author} {\bibfnamefont
  {F.}~\bibnamefont {Zummack}}, \ and\ \bibinfo {author} {\bibfnamefont
  {C.}~\bibnamefont {Sydlo}},\ }\bibfield  {title} {{\selectlanguage
  {english}\enquote {\bibinfo {title} {{F}ew{-F}emtosecond {F}acility{-W}ide
  {S}ynchronization of the {E}uropean {XFEL}},}\ }}in\ \href {\doibase
  doi:10.18429/JACoW-FEL2019-WEB04} {{\selectlanguage {english}\emph {\bibinfo
  {booktitle} {Proc. FEL'19}}}},\ \bibinfo {series and number} {\bibinfo
  {series} {Free Electron Laser Conference}\ No.~\bibinfo {number} {39}}\
  (\bibinfo  {publisher} {JACoW Publishing, Geneva, Switzerland},\ \bibinfo
  {year} {2019})\ pp.\ \bibinfo {pages} {318--321},\ \bibinfo {note}
  {https://doi.org/10.18429/JACoW-FEL2019-WEB04}\BibitemShut {NoStop}%
\bibitem [{\citenamefont {Galayda}(2018)}]{LCLS2}%
  \BibitemOpen
  \bibfield  {author} {\bibinfo {author} {\bibfnamefont {J.}~\bibnamefont
  {Galayda}},\ }\bibfield  {title} {{\selectlanguage {english}\enquote
  {\bibinfo {title} {{T}he {LCLS-II}: {A} {H}igh {P}ower {U}pgrade to the
  {LCLS}},}\ }}in\ \href {\doibase doi:10.18429/JACoW-IPAC2018-MOYGB2}
  {{\selectlanguage {english}\emph {\bibinfo {booktitle} {Proc. of
  IPAC'18}}}},\ \bibinfo {series and number} {\bibinfo {series} {International
  Particle Accelerator Conference}\ No.~\bibinfo {number} {9}}\ (\bibinfo
  {publisher} {JACoW Publishing},\ \bibinfo {address} {Geneva, Switzerland},\
  \bibinfo {year} {2018})\ pp.\ \bibinfo {pages} {18--23},\ \bibinfo {note}
  {https://doi.org/10.18429/JACoW-IPAC2018-MOYGB2}\BibitemShut {NoStop}%
\bibitem [{\citenamefont {Yan}\ and\ \citenamefont {Deng}(2019)}]{shine}%
  \BibitemOpen
  \bibfield  {author} {\bibinfo {author} {\bibfnamefont {J.}~\bibnamefont
  {Yan}}\ and\ \bibinfo {author} {\bibfnamefont {H.}~\bibnamefont {Deng}},\
  }\bibfield  {title} {\enquote {\bibinfo {title} {Multi-beam-energy operation
  for the continuous-wave x-ray free electron laser},}\ }\href {\doibase
  10.1103/PhysRevAccelBeams.22.090701} {\bibfield  {journal} {\bibinfo
  {journal} {Phys. Rev. Accel. Beams}\ }\textbf {\bibinfo {volume} {22}},\
  \bibinfo {pages} {090701} (\bibinfo {year} {2019})}\BibitemShut {NoStop}%
\end{thebibliography}%

\end{document}